\documentclass[sigconf]{acmart}
\usepackage{url}

\usepackage{nicefrac}
\usepackage{siunitx}
\usepackage{array,framed}
\usepackage{pdfpages}
\usepackage{booktabs}
\usepackage{
  color,
  float,
  epsfig,
  wrapfig,
  graphics,
  graphicx,
  subcaption
}
\usepackage{hyperref}
\usepackage{setspace}
\usepackage{amsfonts}
\usepackage{latexsym,fancyhdr}
\usepackage{enumerate}
\usepackage[linesnumbered,ruled,longend,noend]{algorithm2e}
\usepackage{algpseudocode}
\usepackage{graphicx}
\usepackage{xparse} 
\usepackage{xspace}
\usepackage{multirow}
\usepackage{csvsimple}
\usepackage{balance}
\usepackage{bbding}
\usepackage{flushend}
\usepackage{pifont}
\usepackage{threeparttable}
\usepackage{longtable}
\usepackage{wrapfig}
\usepackage{shortcuts}
\usepackage{adjustbox}
\usepackage{booktabs} 
\usepackage{soul}
\usepackage{todonotes}
\usepackage{enumitem}
\usepackage{array}
\usepackage{color}
\usepackage{mdframed}
\usepackage{hyperref}
\usepackage{stfloats}

\AtBeginDocument{%
  }

\begin{document}

\title{\tool{}: Metamorphic Testing for Detecting False Vector Matching Problems in LLM Augmented Generation}

\author{Guanyu Wang}
\email{guanyu.wang@bupt.edu.cn}
\affiliation{
    \institution{Beijing University of Posts and Telecommunications}
}

\author{Yuekang Li}
\email{yuekang.li@unsw.edu.au}
\affiliation{
    \institution{University of New South Wales}
}

\author{Yi Liu}
\email{yi009@e.ntu.edu.sg}
\affiliation{
    \institution{Nanyang Technological University}
}

\author{Gelei Deng}
\email{gdeng003@e.ntu.edu.sg}
\affiliation{
    \institution{Nanyang Technological University}
}

\author{Tianlin Li}
\email{TIANLIN001@e.ntu.edu.sg}
\affiliation{
    \institution{Nanyang Technological University}
}

\author{Guosheng Xu}
\email{guoshengxu@bupt.edu.cn}
\affiliation{
    \institution{Beijing University of Posts and Telecommunications}
}

\author{Yang Liu}
\email{yangliu@ntu.edu.sg}
\affiliation{
    \institution{Nanyang Technological University}
}

\author{Haoyu Wang}
\email{haoyuwang@hust.edu.cn}
\affiliation{
    \institution{Huazhong University of Science and Technology}
}

\author{Kailong Wang}
\email{wangkl@hust.edu.cn}
\affiliation{
    \institution{Huazhong University of Science and Technology}
}

\begin{abstract}

Augmented generation techniques such as Retrieval-Augmented Generation (RAG) and Cache-Augmented Generation (CAG) have revolutionized the field by enhancing large language model~(LLM) outputs with external knowledge and cached information. However, the integration of vector databases, which serve as a backbone for these augmentations, introduces critical challenges, particularly in ensuring accurate vector matching. False vector matching in these databases can significantly compromise the integrity and reliability of LLM outputs, leading to misinformation or erroneous responses. Despite the crucial impact of these issues, there is a notable research gap in methods to effectively detect and address false vector matches in LLM-augmented generation.

This paper presents \tool, a metamorphic testing framework developed to identify false vector matching in LLM-augmented generation systems. We derive eight metamorphic relations (MRs) from six NLP datasets, which form our method's core, based on the idea that semantically similar texts should match and dissimilar ones should not. \tool uses these MRs to create sentence triplets for testing, simulating real-world LLM scenarios. Our evaluation of \tool over 203 vector matching configurations, involving 29 embedding models and 7 distance metrics, uncovers significant inaccuracies. The results, showing a maximum accuracy of only 41.51\% on our tests compared to the original datasets, emphasize the widespread issue of false matches in vector matching methods and the critical need for effective detection and mitigation in LLM-augmented applications.


\end{abstract}

\maketitle

\section{Introduction}
\label{sec:intro}
\begin{figure}[t]
    \centering
    \begin{subfigure}{0.75\linewidth}
        \centering
        \includegraphics[width=1\linewidth]{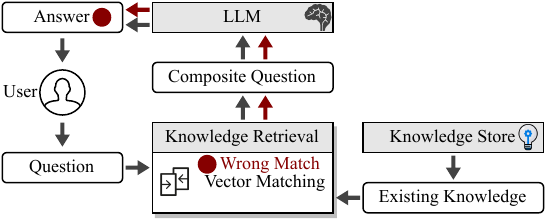}
        \caption{Knowledge Store Type of Application}
        \label{fig:knowledge-app}
    \end{subfigure}
    \centering
    \begin{subfigure}{0.75\linewidth}
        \centering
        \includegraphics[width=1\linewidth]{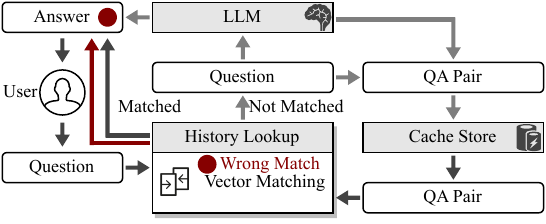}
        \caption{Cache Store Type of Application}
        \label{fig:cache-app}
    \end{subfigure}
    \captionsetup{skip=10pt}\vspace{-0.3cm}
    \caption{Workflow for two types of LLM-based QA applications.}\vspace{-0.6cm}
    \label{fig:app-workflow}
\end{figure}

The landscape of large language models (LLMs) has been profoundly reshaped by the advent of augmented generation techniques, particularly Retrieval-Augmented Generation (RAG) and Cache-Augmented Generation (CAG). These advancements have leveraged the capabilities of vector databases, specialized for storing and retrieving vector representations from embedding models, to significantly enhance LLM functionalities. In \autoref{fig:app-workflow}, systems such as the Cache Store (e.g., GPTCache~\cite{gptcache}) and the Knowledge Store (e.g., Langchain~\cite{langchain}) exemplify this integration, showcasing how these databases bolster LLMs. By efficiently caching and indexing crucial information, they not only streamline the retrieval process but also empower LLMs to deliver more accurate, current, and contextually relevant responses.

The incorporation of vector databases into LLM frameworks holds immense potential for improving information retrieval efficiency. However, it also introduces prominent challenges, particularly in accurate vector matching. The implications of these challenges manifest as two primary issues: miss matching and false matching. Miss matching leads to a fallback on standard information retrieval methods, thus forgoing the efficiency of integrated vector databases. More critically, false matching—akin to a web cache erroneously delivering an unrelated webpage—can significantly impair LLM functionality, either by yielding irrelevant query results (in Cache Store contexts) or by selecting inappropriate background information (in Knowledge Store contexts), potentially leading to misleading or fabricated model outputs.

Despite the critical need to accurately identify false matching for optimal LLM application functionality, current testing methodologies for vector matching methods remain underdeveloped. This gap is further compounded by the absence of a systematic benchmark for evaluating these methods, a challenge exacerbated by the lack of reliable test oracles.

Addressing this research gap, our paper introduces \tool{}, a metamorphic testing framework specifically designed to detect false matching problems in LLM-augmented generation. We derive eight metamorphic relations from six NLP task datasets, categorized into Word-Level and Sentence-Level relations based on metamorphosis degree. \tool{} leverages these relations to structure test cases as triplets: a base sentence, a semantically similar positive sentence with a different structure, and a negative sentence with an inverse meaning. These triplets, generated using AI models, are then used to simulate and test the vector matching methods in LLM-augmented generation applications.

Our comprehensive evaluation of \tool{} spans 203 vector matching configurations, combining 29 embedding models and 7 distance metrics. The results are revealing: a significant decrease in accuracy on our generated test cases compared to original datasets, with the highest accuracy being only 41.51\%. We further conduct a case study on three real-world open source vector databases~\cite{annoy, chroma, scann}, and show that \tool can detect false matching problems in LLM-augmented generation systems. The highest accuracy for correct vector matching is only 37.72\%. This underscores a widespread and non-negligible issue of false matches across all evaluated vector matching methods and highlights the structural bias over semantic accuracy in these systems. 

\textbf{Contribution.} The key contributions are enumerated as follows:




\begin{itemize}
\item \textbf{Innovative Metamorphic Testing Framework for LLM-Augmented Generation.} We introduce a groundbreaking framework specifically for evaluating vector matching in LLM-augmented generation, marking a first in the field to the best of our knowledge. A key contribution is the development of eight semantic-based metamorphic relations at the sentence level. These relations are crucial for benchmarking LLM-augmentation techniques.


\item \textbf{Comprehensive Evaluation of Matching Methods.} Our study conducts a thorough evaluation of 203 vector matching methods, combining various embedding models and distance metrics. This extensive analysis offers crucial insights into the performance and limitations of current matching techniques in LLM contexts.

\item \textbf{Insights into False Matching in Augmented LLMs.} A significant finding is the tendency of vector matching methods to prioritize structural over semantic similarities, contrasting with the more effective semantic discernment of text matching methods. This highlights areas for improvement in matching techniques in LLM-augmented generation.
\end{itemize}

We have open-sourced the dataset and more detailed experiment results at ~\cite{ourtool} to facilitate open-science and future research.



\section{Background} 
\label{sec:background}

\subsection{Vector Databases}
A vector database is a specialized type of database designed to handle vector data, which are essentially multi-dimensional arrays or lists of numerical values often used to represent features in machine learning models, images, or any other data with multiple attributes. Unlike traditional databases that handle scalar values like integers or strings, vector databases are optimized for high-performance similarity search, allowing users to quickly retrieve the most similar vectors to a given query vector based on various distance or similarity metrics like Cosine Similarity or Euclidean Distance.

Key features of a vector database often include scalability, low-latency retrieval, and support for multiple similarity metrics. They are also designed to handle massive datasets and can be distributed across multiple nodes to improve search performance. Advanced vector databases may also offer features like auto-indexing, query-time filtering, and integration with machine learning frameworks, thus making them particularly useful in applications ranging from recommendation systems to computer vision and natural language processing. The major concern, however, is the accuracy of the information retrieval, which is the key focus of this work.

\vspace{-0.3cm}
\subsection{Metamorphic Testing}
Metamorphic testing~\cite{ChenTY2018} serves as a dual-purpose technique for both generating test cases and verifying test results, strategically addressing the so-called ``oracle problem'' in many software testing scenarios~\cite{Wang_ICSE23,Xiao2022MetamorphicTO,Yuan2022UnveilingHD,Chen2020MetamorphicTA,Ma2020MetamorphicTA,Santos2020AnES}. 
At the core of this methodology lies the concept of metamorphic relations (MRs), key properties that delineate the expected associations between sets of input-output pairs in the software being tested.
The procedure entails taking an initial test case and applying a predefined transformation rule to produce a closely related new test case. 
Subsequently, the outputs of these paired test cases are scrutinized to confirm their adherence to the established MRs. Since its inception, the domain of metamorphic testing has witnessed substantial scholarly activity, spanning a range of focus areas including the identification of MRs, innovative test case generation strategies, integration with existing software engineering paradigms, and rigorous validation and evaluation of software systems.
In particular, we propose MRs for textual contents to test the semantic similarity for queries in LLM-integrated applications~(\autoref{fig:app-workflow}) in this work. 
\section{Methodology}
\label{sec:methodology}

\begin{figure*}[t]
\includegraphics[width=1\textwidth]{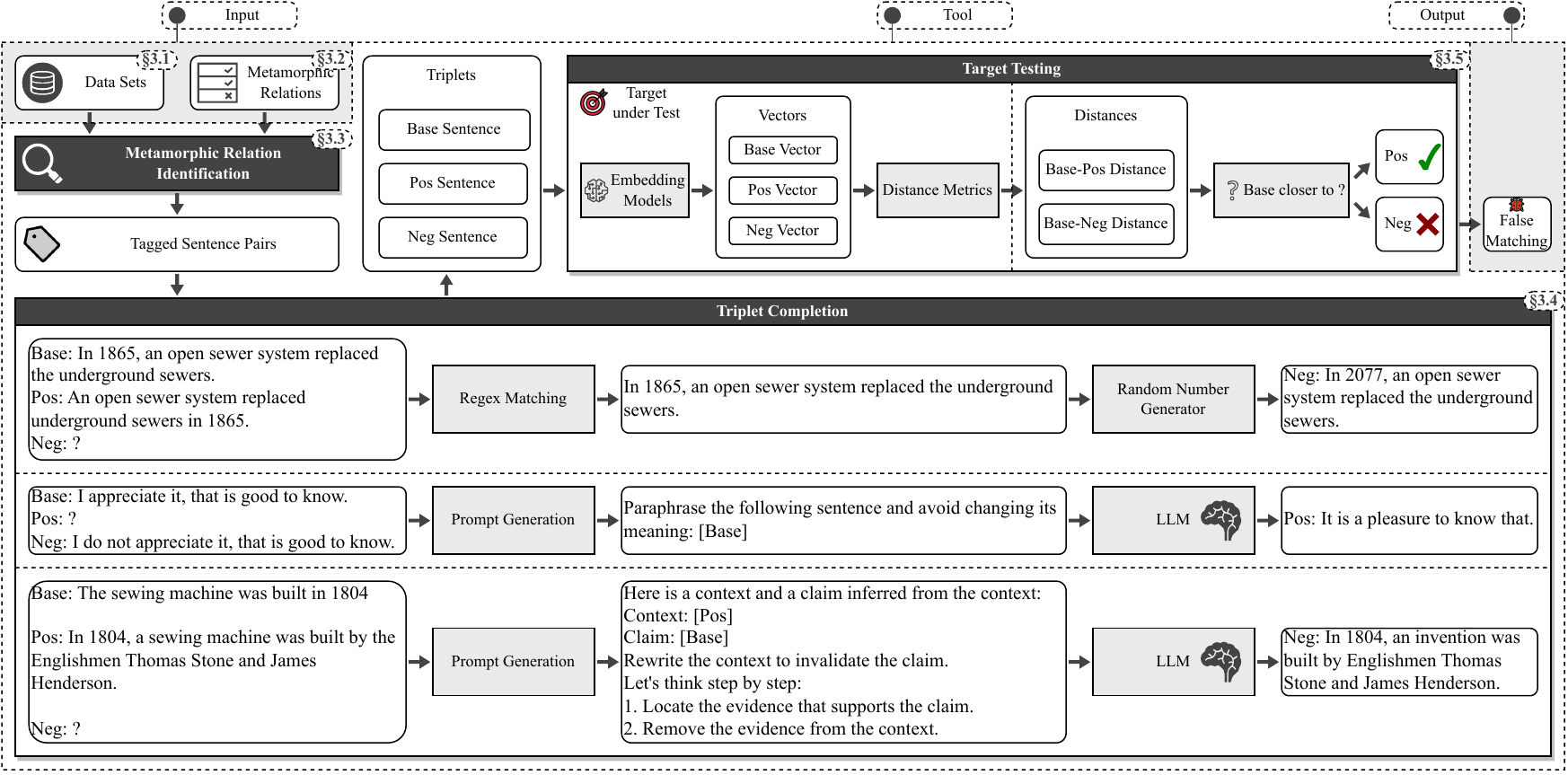}\vspace{-0.3cm}
\caption{The overview of \tool. Includes three parts: sentence pairs collection, triplet complementation, and simulation of vector matching.}\vspace{-0.3cm}
\label{fig:overview}
\end{figure*}

In this section, we introduce our \tool. The overview of our testing framework is shown in \autoref{fig:overview}. We aim to construct a corpus where each sample is a triplet of the following form:
\begin{align}
    \{Query:Base\ |\ Candidates:Positive, Negative\}\label{eq:triplet}
\end{align}
Semantically, the base sentence and the positive are considered identical sentences, but contradictory to the negative. In terms of sentence structure, the structural similarity between base and negative sentences is higher than with positive sentences. We first conduct a pilot study to look for metamorphosis that may provoke changes at the semantic level (\autoref{sec:pilot study}). Then, we introduce eight MRs from the pilot study in \autoref{sec:metamorphosis} and collect sentence pairs from six existing datasets in \autoref{sec:filter}. We complement the sentence pairs into triplets in \autoref{sec:generator} and show how to use them for testing in \autoref{sec:retrieval}.


\subsection{Pilot Study}
\label{sec:pilot study}

In this work, we intend to explore how to make semantic changes with as few changes to a sentence as possible. To design sufficiently effective metamorphosis, we first conduct a pilot study on existing datasets to explore which metamorphosis meet our expectation. We consider the following 6 sentence relation datasets, including \textbf{Stanford Contradiction Corpora~\cite{de2008finding}}, \textbf{PAWS~\cite{zhang2019paws}}, \textbf{VitaminC~\cite{schuster2021get}}, 
\textbf{Inference-is-Everything~\cite{white2017inference}},
\textbf{HEROS~\cite{chiang2023revealing}} and \textbf{NEVIR~\cite{weller2023nevir}}.

To answer our question, we collect samples labelled as contradictory etc. indicating non-consistent relationships between sentence pairs. Then, we analyse the structure of these samples. Specifically, We collect 500 sentences pairs from each of the datasets. We iterate through each pairs, looking for differences in the sentences. We only consider samples that contain more identical words. This is because in such cases, sentence pairs are more likely generated by some kind of morphing rather than by a random combination of two unrelated sentences. We categorise the metamorphic methods into word-level and sentence-level based on the extent of the metamorphosis. Word-level includes Word Swap-based, Object Substitution-based, Action Substitution-based, Negative Expression-based, Word Deletion, Quantifier Substitution-based. Their metamorphosis occurs only on a particular word or phrase. Sentence-level includes Back Translation-based and Inference Generation-based. They usually produce more changes in sentence structure. We summarise eight MRs accordingly~(to be detailed in Section~\ref{sec:metamorphosis}).

\vspace{-0.3cm}
\subsection{Metamorphic Relation}
\label{sec:metamorphosis}

In this section, we introduce eight MRs inspired by the pilot study. To facilitate the unambiguous understanding, we formally define them. We further classify them into two types based on the degree of metamorphosis: \textbf{Word-Level} metamorphosis and \textbf{Sentence-Level} metamorphosis. In particular, we first define the base sentence, from which we derive the negative sentence using metamorphic transformation~(\autoref{sec:filter}). Then, the positive sentences are completed by the generator~(\autoref{sec:generator}).

\vspace{-0.3cm}
\subsubsection{\textbf{Word-Level Metamorphosis}}
\label{Word-Level Metamorphosis}

\begin{table*}[t]
\caption{Metamorphic Relations}\vspace{-0.3cm}
\label{table:relation}
\begin{adjustbox}{width=1\textwidth}
\begin{tabular}{ccp{11cm}l}
\toprule
\textbf{Metamorphic Levels}                  & \textbf{Relation Names/Shortcuts}                                                                                  & \multicolumn{1}{c}{\textbf{Examples}}                                                                                                                                       & \multicolumn{1}{c}{\textbf{Sources}} \\ \midrule
\multirow{21}{*}{\textbf{Word Level}}    & \multirow{3}{*}{\textbf{\begin{tabular}[c]{@{}c@{}}Word Swap\\ (Word Swap, WS)\end{tabular}}}               & \textbf{\textit{Base:}} The \textcolor{red}{only} industry in the town is \textcolor{red}{light} farming on the small rice paddies.                                                                                            & \textbf{\textit{Base:}} Collect                       \\
                                         &                                                                                                         & \textbf{\textit{Posi:}} Light farming on the small rice paddies is the only industry in the town.                                                                                            & \textbf{\textit{Posi:}} Generate                      \\
                                         &                                                                                                         & \textbf{\textit{Nega:}} The \textcolor{red}{light} industry in the town is \textcolor{red}{only} farming on the small rice paddies.                                                                                            & \textbf{\textit{Nega:}} Collect                       \\ \cline{2-4} 
                                         & \multirow{3}{*}{\textbf{\begin{tabular}[c]{@{}c@{}}Object Substitution\\ (Obj Sub, OS)\end{tabular}}}       & \textbf{\textit{Base:}} Carl borrowed a book from Richard, but the book was unreadable to \textcolor{red}{him}.                                                                                               & \textbf{\textit{Base:}} Collect                       \\
                                         &                                                                                                         & \textbf{\textit{Posi:}} Carl was unable to make sense of the book he borrowed from Richard.                                                                                                  & \textbf{\textit{Posi:}} Generate                      \\
                                         &                                                                                                         & \textbf{\textit{Nega:}} Carl borrowed a book from Richard , but the book was unreadable to \textcolor{red}{Richard}.                                                                                          & \textbf{\textit{Nega:}} Collect                       \\ \cline{2-4} 
                                         & \multirow{3}{*}{\textbf{\begin{tabular}[c]{@{}c@{}}Action Substitution\\ (Act Sub, AS)\end{tabular}}}       & \textbf{\textit{Base:}} I think we \textcolor{red}{know} what we 're going to speak about.                                                                                                                    & \textbf{\textit{Base:}} Collect                       \\
                                         &                                                                                                         & \textbf{\textit{Posi:}} I believe we are aware of what to discuss.                                                                                                                           & \textbf{\textit{Posi:}} Generate                      \\
                                         &                                                                                                         & \textbf{\textit{Nega:}} I think we \textcolor{red}{be} what we 're going to speak about.                                                                                                                      & \textbf{\textit{Nega:}} Collect                       \\ \cline{2-4} 
                                         & \multirow{6}{*}{\textbf{\begin{tabular}[c]{@{}c@{}}Negative Expression\\ (Nega Exp, NE)\end{tabular}}}      & \textbf{\textit{Base:}} I appreciate it, that is good to know.                                                                                                                               & \multirow{2}{*}{\textbf{\textit{Base:}} Collect}      \\
                                         &                                                                                                         & \textbf{\textit{Posi:}} It is a pleasure to know that.                                                                                                                                       &                                     \\
                                         &                                                                                                         & \textbf{\textit{Nega:}} I \textcolor{red}{do not} appreciate it, that is good to know.                                                                                                                        & \multirow{2}{*}{\textbf{\textit{Posi:}} Generate}     \\
                                         &                                                                                                         & \textbf{\textit{Base:}} But it is Jackie saying it so sorta \textcolor{red}{disappointing}.                                                                                                                   &                                     \\
                                         &                                                                                                         & \textbf{\textit{Posi:}} Jackie saying it makes it somewhat disappointing.                                                                                                                    & \multirow{2}{*}{\textbf{\textit{Nega:}} Collect}      \\
                                         &                                                                                                         & \textbf{\textit{Nega:}} But it is Jackie saying it so sorta \textcolor{red}{upbeat}.                                                                                                                          &                                     \\ \cline{2-4} 
                                         & \multirow{3}{*}{\textbf{\begin{tabular}[c]{@{}c@{}}Word Deletion\\ (Word Del, WD)\end{tabular}}}            & \textbf{\textit{Base:}} In 2012, Jordan started all 16 games while recording 8.0 sacks and 54 tackles.                                                                                       & \textbf{\textit{Base:}} Collect                       \\
                                         &                                                                                                         & \textbf{\textit{Posi:}} In 2012, jordan started 16 games and recorded 8.0 sacks and 54 tackles.                                                                                              & \textbf{\textit{Posi:}} Generate                      \\
                                         &                                                                                                         & \textbf{\textit{Nega:}} \textcolor{red}{\st{In 2012}}, Jordan started all 16 games while recording 8.0 sacks and 54 tackles.                                                                                       & \textbf{\textit{Nega:}} Collect                       \\ \cline{2-4} 
                                         & \multirow{3}{*}{\textbf{\begin{tabular}[c]{@{}c@{}}Quantifier Substitution\\ (Quant Sub, QS)\end{tabular}}} & \textbf{\textit{Base:}} In \textcolor{red}{1865}, an open sewer system replaced the underground sewers.                                                                                                       & \textbf{\textit{Base:}} Collect                       \\
                                         &                                                                                                         & \textbf{\textit{Posi:}} An open sewer system replaced underground sewers in 1865.                                                                                                            & \textbf{\textit{Posi:}} Generate                      \\
                                         &                                                                                                         & \textbf{\textit{Nega:}} In \textcolor{red}{3016} , an open sewer system replaced the underground sewers.                                                                                                      & \textbf{\textit{Nega:}} Generate                      \\ \hline
\multirow{6}{*}{\textbf{Sentence Level}} & \multirow{3}{*}{\textbf{\begin{tabular}[c]{@{}c@{}}Back Translation\\ (Err Trans, ET)\end{tabular}}}        & \textbf{\textit{Base:}} The second series was well received by the critics better than the first.                                                                                            & \textbf{\textit{Base:}} Collect                       \\
                                         &                                                                                                         & \textbf{\textit{Posi:}} The critics received the second series more favorably than the first.                                                                                                & \textbf{\textit{Posi:}} Generate                      \\
                                         &                                                                                                         & \textbf{\textit{Nega:}} The first series was recorded by critics better than the second.                                                                                                     & \textbf{\textit{Nega:}} Collect                       \\ \cline{2-4} 
                                         & \multirow{3}{*}{\textbf{\begin{tabular}[c]{@{}c@{}}Inference Generation\\ (Err Nli, EN)\end{tabular}}}        & \textbf{\textit{Base:}} The sewing machine was built in 1804.                                                                                                                                & \textbf{\textit{Base:}} Collect                       \\
                                         &                                                                                                         & \textbf{\textit{Posi:}} In 1804 , a sewing machine was built by the Englishmen Thomas Stone and James Henderson , and a machine for embroidering was constructed by John Duncan in Scotland. & \textbf{\textit{Posi:}} Collect                       \\
                                         &                                                                                                         & \textbf{\textit{Nega:}} In 1804, an invention was built by Englishmen Thomas Stone and James Henderson, and a device for embroidering was constructed by John Duncan in Scotland.            & \textbf{\textit{Nega:}} Generate                      \\ \bottomrule
\end{tabular}
\end{adjustbox}\vspace{-0.3cm}
\end{table*}

In this level, we choose sentence pairs that result in semantic contradictions by deleting, replacing, or swapping the order of the words. The samples at this level are characterized with very similar sentence structures, and only individual words are metamorphosed. Specifically, given a base sentence, we aim to derive the corresponding negative sentence using the MR. 

\begin{definition}
  \textbf{(Base Sentence)} The base sentence can be defined by an N-tuple: 
  \begin{center}
  Base sentence = $<w_1, w_2, ..., w_t, ..., w_{n-1}, w_n>$,
  \end{center}
  where $w_k (0 \le k \le n)$ represents a constituent word in the sentence, $w_t$ represent a target word.
\end{definition}

\textbf{Word Swap.} This MR yields a contradiction by reordering the words. 

\begin{definition}
  \textbf{(Word Swap)} Given the base sentence:
  \begin{center}
  Base sentence = $<w_1, w_2, ..., w_t, ..., w_{t+j}, ..., w_{n-1}, w_n>$, 
  \end{center}
   where $w_t$ and $w_{t+j}$ are two target words. Then we can obtain the negative sentence:
  \begin{center}
  Negative sentence = $<w_1, w_2, ..., w_{t+j}, ..., w_{t}, ...w_{n-1}, w_n>$,
  \end{center}
\end{definition}

\textbf{Object Substitution.} This MR replaces a word or phrase in a sentence that can refer to an object to create an inconsistency, which are usually nouns or pronouns. 

\begin{definition}
  \textbf{(Object Substitution)} Given the target word $w_t$ being an object, the negative sentence is given:
  \begin{center}
  Negative sentence = $<w_1, w_2, ..., w_{obj}, ..., w_{n-1}, w_n>$,
  \end{center}
  where $w_{obj}$ represents a different object compared with the original target word.
\end{definition}

\textbf{Action Substitution.} This MR replaces an action described in a sentence to create a contradiction. 

\begin{definition}
  \textbf{(Action Substitution)} Given the target word $w_t$ being a verb, the negative sentence is given:
  \begin{center}
  Negative sentence = $<w_1, w_2, ..., w_{verb}, ..., w_{n-1}, w_n>$,
  \end{center}
  where $w_{verb}$ represents a different action represented by the original target word.
\end{definition}

\textbf{Negative Expression.} This MR adds negatives to the front of the verb or uses antonyms to replace adjectives in base sentence to produce semantic contradictions. 

\begin{definition}
  \textbf{(Negative Substitution)} Given the target word $w_t$ being a verb, adverb or adjective, the negative sentence is given:
  \begin{center}
  Negative sentence = $<w_1, w_2, ..., w_{neg}, ..., w_{n-1}, w_n>$,
  \end{center}
  where $w_{neg}$ represents one or a set of semantically negative word(s) versus the original target word.
\end{definition}

\textbf{Word Deletion.} This MR deletes words or phrases in base sentence to generate contradictions. One point worth noting is that this category of samples is not completely dissimilar to the other categories. The reason is that deleting a part of a sentence may amplify the range of meanings expressed in the sentence. This means that it can infer metamorphic sentences from original sentences, but not feasible in the opposite direction~(as listed in \autoref{table:relation}, we can infer the negative sentence from the base sentence, but in turn we can't infer which year the events described in the negative sentence happen in). We denote this phenomenon as \textit{One-Way Inconsistency}, which will bring a new challenge to semantic evaluation. 

\begin{definition}
  \textbf{(Word Deletion)} Given the target word $w_t$, the negative sentence is given:
  \begin{center}
  Negative sentence = $<w_1, w_2, ..., remove(w_{t}), ..., w_{n-1}, w_n>$,
  \end{center}
  where $remove(w_{t})$ represents the deletion of the original target word.
\end{definition}

\textbf{Quantifier Substitution.} This MR generates contradiction by identifying and changing the quantifier in a sentence. 

\begin{definition}
  \textbf{(Quantifier Substitution)} Given the target word $w_t$ being a quantifier, the negative sentence is given:
  \begin{center}
  Negative sentence = $<w_1, w_2, ..., w_{quantifier}, ..., w_{n-1}, w_n>$,
  \end{center}
  where $w_{quantifier}$ represents a different quantifier compared with the original target word.
\end{definition}

\subsubsection{\textbf{Sentence-Level Metamorphosis}}
\label{Sentence-Level Metamorphosis}

In this level, the metamorphosis of the original sentence is no longer limited to a particular word or phrase. That means base and negative sentences have not only different meanings but even more different sentence structures in this level. Due to the diverse patterns and possibilities for generating the triplet, we utilize informal definition to describe the MRs.  

\textbf{Back Translation.} This MR records changes in sentence meaning due to back translation errors. It can be seen as a free combination of \textbf{Word-Level} MRs. 
    
\textbf{Inference generation.} This MR produces contradictory sentence pairs through incorrect NLI. 
Similarly, \textit{One-Way Inconsistency} exists in this category. That is, it can infer information about the base sentence from the positive sentence, but the inference doesn't stand up in the opposite direction.

\subsection{MR Identification}
\label{sec:filter}

\SetKwFunction{FindDiff}{FindDiff}
\SetKwFunction{NLTK}{NLTK}
\SetKwFunction{Regex}{Regex}
\SetKwFunction{GoldTag}{GoldTag}
\SetKwProg{Fn}{Function}{:}{}

\begin{algorithm}
\scriptsize
\caption{Word-Level Metamorphosis Tagging}\label{alg:filter}
\KwData{Sentence Pair $<S1,S2>$ and $label$ from the existing datasets}
\KwResult{$Tag$}
\Fn{\GoldTag{$<S1,S2>$, $label$}}{
\eIf{$label$ \textbf{is "Entailment"}}{
    $Tag \gets "Other"$
}{
    $W1 \gets S1.Lower.Split$\;
    $W2 \gets S2.Lower.Split$\;
    \eIf{$W1.Sort = W2.Sort$}{
    \tcp{Sorting the word lists.}
        $Tag \gets "WordSwap"$
    }{
        $U \gets W1\cup W2$\;
        \eIf{$|W1| = |W2|$ \textbf{and} ($|U| - |W1| \leq 1$ \textbf{or} $|U| - |W2| \leq 1$)}{
            $diff_1,diff_2 \gets \FindDiff{W1,W2}$\; 
            \eIf{$diff_1=Regex(S1)$ \textbf{and} $diff_2=Regex(S2)$}{\tcp{Regex match quantifiers}
                $Tag \gets "QuantSub"$
            }{
                $tag\_dict1, tag\_dict2 \gets NLTK(S1, S2) $\;
                $tag1 \gets tag\_dict1[diff_1]$\;
                $tag2 \gets tag\_dict2[diff_2]$\;
                \eIf{$tag1 \subset [NN, PRP, JJ, RB, VB]$ \textbf{and} $tag2 \subset [NN, PRP, JJ, RB, VB]$}{ \tcp{Identify the part of speech. NN and similar terms represent word parts of speech.}
                    \lIf{$tag1 \subset [NN, PRP]$ \textbf{and} $tag2 \subset [NN, PRP]$}{
                        $Tag \gets "ObjSub"$
                    }
                    \lElseIf{$tag1 \subset [JJ, RB]$ \textbf{and} $tag2 \subset [JJ, RB]$}{
                        $Tag \gets "NegaExp"$
                    }
                    \lElseIf{$tag1$ \textbf{is "VB"} \textbf{and} $tag2$ \textbf{is "VB"}}{
                        $Tag \gets "ActSub"$
                    }
                    \lElse{
                        $Tag \gets "Other"$
                    }
                }{
                    $Tag \gets "Other"$
                }
            }
        }{
        \eIf{$W1 \subset W2$}{ \tcp{Assuming $|W1| < |W2|$}
            \eIf{$|W2| - |W1| \leq 2$ \textbf{and "not" in} $W2-W1$}{
                $Tag \gets "NegaExp"$
            }{
                $Tag \gets "WordDel"$
            }
        }{
            $Tag \gets "Other"$
        }
        }
    }
\KwRet $Tag$
}
}
\pagebreak
\Fn{\FindDiff{$W1$, $W2$}}{
    \For{$w1,w2$ \textbf{in} $W1,W2$}{
    \If{$w1 \neq w2$}{
    \KwRet $w1,w2$
    }
    }
}
\Fn{\Regex{$Sentence$}}{
    $quant \gets findall("-?\textbackslash d+(?:\textbackslash .\textbackslash d+)?")$\;
    \KwRet $quant$
}
\Fn{\NLTK{$S1, S2, diff_1, diff_2$}}{
    $tag\_dict1 \gets nltk.word\_tokenize(S1)$\;
    $tag\_dict2 \gets nltk.word\_tokenize(S2)$\;
    \KwRet $tag\_dict1, tag\_dict2$
}
\end{algorithm}


In this section, we show given a sentence pair, how to determine whether it matches our MRs and with which MR it matches.

We use the MR Identification algorithm to distinguish Word-Level MRs. As shown in \autoref{alg:filter}: Given a sentence pair and its relation label, we first determine whether they are in an inconsistent relation (line 2,3). Then we convert both sentences to lower case to eliminate case effects and split them into two sets of words, W1 and W2. (line 5,6). When the sorting of W1 and W2 is consistent, it indicates that the inconsistency in the sentence pairs resulted from swapping of words (line 7,8). If two sentences are the same length and there are inconsistencies due to word substitution, we first traverse W1 and W2 to determine which word has been substituted (line 11,12). Once the word is found, we first use a regular expression to determine if the metamorphosis word is a Quantifier, and if it is, the word is marked as a Quant Sub (line 13,14). Otherwise, we use NLTK~\cite{loper2002nltk} to judge the components of the word in the original sentence (line 15-25). Our focus will be on five categories: nouns, pronouns, adjectives, adverbs and verbs. If both are nouns or pronouns, label it as Obj Sub. If they are all adjectives or adverbs, mark it as NegaExp. If they are both verbs, tag it as Obj Sub. Otherwise, it is assumed that we no longer need to consider it. When two sentences are unequal in length and W1 is completely contained within W2 (line 27), and If the missing part of W1 denotes a negation, e.g. not, do not, etc., this can be considered as the addition of a negative word, recorded as Nega Exp (line 29). Otherwise, we assume that this inconsistency was caused by Word Del (line 31).

In identifying sentence-level MR, as present in the previous section, back translation can be viewed as a free combination of word-level MRs. The most obvious feature of these use cases is that the sentences are close in length and the keywords in the sentences are basically the same. Modifiers in one sentence can often be found in another, or in the form of synonyms in another. Therefore, we screen the inconsistencies generated through back translation based on the above features. And in inference generation, the claim can usually be localised to a specific place in the context. In other words, there is a free compositional transformation relation between a claim sentence and the context. We consider this feature to identify pairs of sentences constructed by this relationship and generate negative sentences accordingly.

\vspace{-0.3cm}
\subsection{Triplet Completion}
\label{sec:generator}

We generate the third sentence in triplets based on the two collected sentences. In total, this can be divided into three cases: 

\begin{enumerate}
    \item Generating negative sentences by replacing quantifiers in base sentences.
    
    \item Generating positive sentences by rewriting base sentences.
    
    \item Generating negative sentences by destroying the inference relation between base and positive sentences.
\end{enumerate}

In the case of the former, firstly, we need to recognize and extract the quantifiers in the sentence. This can be accomplished by employing a regular expression, which is a powerful text matching tool used to locate text fragments that conform to specific patterns. In this context, it's used to match quantifiers. Once we've successfully extracted the quantity from the sentence, the next step is to randomly adjust it. This means multiplying the original quantity by a random seed within the range of 0 to 2, excluding 1. Finally, we substitute the newly generated random number in place of the initial quantity within the sentence. This replacement effectively changes the sentence's meaning by altering the quantified value.

In the second case, we opt for the Text-Davinci-003 model. This model is well-suited for text generation tasks and is capable of comprehending and manipulating text effectively. We construct prompt word to achieve the desired transformation of the sentence. The prompt give the model hints and guidance on how to rewrite the sentence. They are designed to influence the model's output while keeping the core meaning intact. By employing this method, we can generate affirmative sentences that exhibit structural variations while retaining the original statement.

For the last case, we use the Davinci model for completion as well. More specifically, we construct the prompt words along the Chain-of-Thought~\cite{wei2022chain} style. The purpose of the prompt word is to be a step-by-step guide for the model towards the target task. The process begins by directing the model to identify and focus on specific information pertaining to the supporting evidence that connects the claim to the surrounding context. Subsequently, we instruct the model to systematically remove this evidence-based information from the original context. The intention here is to disrupt the inferential relationship that previously existed. Until then, in this revised context, the inference of the claim would have been considered infeasible.

\vspace{-0.3cm}
\subsection{Simulation of Information Retrieval}
\label{sec:retrieval}

In this section, we introduce how we measure the effectiveness of a vector matching method. 

As shown in \autoref{fig:overview}, in this process, the first step involves the individual mapping of each sentence into a high-dimensional feature vector using a uniform embedding model. This mapping process facilitates the subsequent comparison of their similarity within a given feature space. This similarity is quantified using a particular distance metric, such as Euclidean Distance. We create both positive and negative vectors as the potential candidates. Using the base vector as our query vector, we emulate a scenario of retrieving information from a vector database. The goal is to find the item in the database that has the highest similarity to the query vector. We calculate the distances between the base vector and each of the two candidates separately. and the candidate vector with the smaller distance is then selected as the output of this matching operation. If the result point to the positive sentence, we consider the match correct. Conversely, we consider the matching erroneous.
\section{Evaluation } 
\label{sec:eval}

\begin{table*}[t]
\caption{Models and Methods under test and corresponding Markers in the paper}\vspace{-0.3cm}
\label{table:model&mark}
\begin{spacing}{1.3}
\begin{adjustbox}{width=1\textwidth}
\begin{tabular}{cccccc}
\toprule
\textbf{Category}                           & \textbf{Sources/Types}                                                                                      & \multicolumn{4}{c}{\textbf{Models/Methods  \&  Marks}}                                                                              \\ \midrule
\multirow{15}{*}{\textbf{Embedding Models}} & \multirow{5}{*}{\textbf{\begin{tabular}[c]{@{}c@{}}Open Source Projects or \\ Vector Databases\end{tabular}}}                                                              & Cohere/embed-english-v2.0~\cite{cohere}              & \textbf{Cohere-embed}  & Paddle/ernie-3.0-medium-zh~\cite{wang2021ernie}              & \textbf{Paddle-ernie}   \\
                                            &                                                                                                             & FastText-en~\cite{joulin2016fasttext}                            & \textbf{FastText}      & sgugger/rwkv-430M-pile~\cite{rwkv}                  & \textbf{Rwkv}           \\
                                            &                                                                                                             & Distilbert-base-uncased~\cite{Sanh2019DistilBERTAD}                & \textbf{DistilBERT}    & all-MiniLM-L6-v2~\cite{wang2020minilm}                        & \textbf{MiniLM}         \\
                                            &                                                                                                             & Paraphrase-albert-onnx~\cite{albertonnx}                 & \textbf{AlBERT-onnx}   & unum-cloud/uform-vl-english~\cite{uform}             & \textbf{Uform}          \\
                                            &                                                                                                             & text-embedding-ada-002~\cite{neelakantan2022text}                 & \textbf{Ada}           &                                         & \textbf{}               \\ \cline{2-6} 
                                            & \multirow{5}{*}{\textbf{\begin{tabular}[c]{@{}c@{}}SBERT's List or \\ HuggingFace Model Hub\end{tabular}}} & bert-base-uncased~\cite{DBLP:journals/corr/abs-1810-04805}                      & \textbf{BERT}          & SpanBERT/spanbert-large-cased~\cite{joshi2020spanbert}           & \textbf{SpanBERT}       \\
                                            &                                                                                                             & michiyasunaga/LinkBERT-base~\cite{yasunaga2022linkbert}            & \textbf{LinkBERT}      & google/electra-large-generator~\cite{clark2020electra}          & \textbf{Electra}        \\
                                            &                                                                                                             & xlnet-large-cased~\cite{DBLP:journals/corr/abs-1906-08237}                      & \textbf{XLNet}         & sentence-transformers/gtr-t5-large~\cite{ni2021large}      & \textbf{GTR-T5}         \\
                                            &                                                                                                             & roberta-base~\cite{liu2019roberta}                           & \textbf{RoBERTa}       & sentence-transformers/sentence-t5-large~\cite{ni2021sentence} & \textbf{Sentence-T5}    \\
                                            &                                                                                                             & albert-base-v2~\cite{DBLP:journals/corr/abs-1909-11942}                         & \textbf{AlBERT}        & sentence-transformers/all-mpnet-base-v2~\cite{mpnet} & \textbf{MPNet}          \\ \cline{2-6} 
                                            & \multirow{5}{*}{\textbf{The latest LLMs}}                                                                   & tiiuae/falcon-7b                       & \textbf{Falcon-7B}     & tiiuae/falcon-7b-4bit                   & \textbf{Falcon-7B-Q}    \\
                                            &                                                                                                             & decapoda-research/llama-7b-hf          & \textbf{LLaMA-7B}      & decapoda-research/llama-7b-hf-4bit      & \textbf{LLaMA-7B-Q}     \\
                                            &                                                                                                             & decapoda-research/llama-13b-hf         & \textbf{LLaMA-13B}     & decapoda-research/llama-13b-hf-4bit     & \textbf{LLaMA-13B-Q}    \\
                                            &                                                                                                             & meta-llama/Llama-2-7b-hf               & \textbf{LLaMA2-7B}     & meta-llama/Llama-2-7b-hf-4bit           & \textbf{LLaMA2-7B-Q}    \\
                                            &                                                                                                             & meta-llama/Llama-2-13b-hf              & \textbf{LLaMA2-13B}    & meta-llama/Llama-2-13b-hf-4bit          & \textbf{LLaMA2-13B-Q}   \\ \hline
\multirow{4}{*}{\textbf{Distance Metrics}}  & \multirow{4}{*}{\textbf{Scipy}}                                                                             & Cosine Distance                        & \textbf{CD}            & Lance and Williams Distance             & \textbf{LD}             \\
                                            &                                                                                                             & Euclidean Distance                     & \textbf{ED}            & Pearson Correlation Distance            & \textbf{PD}             \\
                                            &                                                                                                             & Manhalanobis Distance                   & \textbf{MD}            & Manhattan Distance                      & \textbf{MhD}            \\
                                            &                                                                                                             & Bray Curtis Distance                   & \textbf{BD}            &                                         & \textbf{}               \\ \hline
\multirow{6}{*}{\textbf{Baselines}}         & \textbf{QG\&QA-based}                                                                                       & QuestEval                              & \textbf{QuestEval}     &                                         & \textbf{}               \\ \cline{2-6} 
                                            & \multirow{3}{*}{\textbf{Cross-encoders}}                                                                    & cross-encoder/quora-distilroberta-base & \textbf{DistilRoBERTa} &                                         & \textbf{}               \\
                                            &                                                                                                             & nli-deberta-v3-base                    & \textbf{DeBERTa}    & nli-deberta-v3-base-reverse             & \textbf{DeBERTa-R}   \\
                                            &                                                                                                             & roberta-large-mnli                     & \textbf{RoBERTa-mnli}  & roberta-large-mnli-reverse              & \textbf{RoBERTa-mnli-R} \\ \cline{2-6} 
                                            & \multirow{2}{*}{\textbf{NLI models-based}}                                                                  & summac-mnli-zs                         & \textbf{SummaC-zs}     & summac-mnli-zs-reverse                  & \textbf{SummaC-zs-R}    \\
                                            &                                                                                                             & summac-mnli-conv                       & \textbf{SummaC-conv}   & summac-mnli-conv-reverse                & \textbf{SummaC-conv-R}  \\ \bottomrule
\end{tabular}
\end{adjustbox}
\end{spacing}\vspace{-0.3cm}
\end{table*}

In this section, we evaluate whether \tool can detect \error. We simulate the process of vector matching in 203 (29 models and 7 distance metrics combined) vector matching methods for validation. We aim to answer the following research questions:
 
\begin{itemize}[itemsep=0pt,parsep=0pt,topsep=0pt,partopsep=0pt]
    \item \textbf{RQ1:} Are these test cases generated by our approach consistent with our requirements?

    \item \textbf{RQ2:} Is \tool capable of identifying instances of false matching in vector databases

    \item \textbf{RQ3:} Can our approach find the erroneous outputs returned by information retrieval methods?

    \item \textbf{RQ4:} How different factors affect the performance of \tool?
\end{itemize}

\vspace{-0.3cm}
\subsection{Experimental Settings} 

\noindent \textbf{Dataset.} 
We build a dataset containing 5,000 triplets for each MRs as data for our experiments. Initially, we gathered 5,000 sentences with quantifiers from the source datasets and generated corresponding positive and negative sentences. Next, we collected 5,000 implication pairs as positive sentences for Err Nli and formulated the negative sentences. For the remaining six MRs, we identified contradiction types in the source datasets to assemble base and negative sentences, generating positive sentences through rewriting. We get a corpus with eight sub-datasets, totalling 40,000 triplets. Furthermore, we modified the benchmark to simulate a non-metamorphic scenario. We maintain the base and positive sentences of the $i$ th and $i+1$ th~(where $i$ is odd) use cases but swap their negatives with the other's positive. In these test cases, both candidates differ structurally from the base sentence, yet only one preserves the same meaning. We conduct our experiments using the two corpora.

\noindent \textbf{Models Under Test.} 
To answer RQ2, we chose three vector databases integrated with Langchain for testing:  Annoy~\cite{annoy}, Chroma~\cite{chroma}, ScaNN~\cite{scann}. For simplifying RQ3's experimental process and encompassing diverse methods, we simulated vector matching. We extracted embedding models used from the open source vector databases integrated with Langchain and projects in Github. For methods that provide a generic interface, we extracted the default model in the method. Additionally, we selected 10 models from SBERT's list and the Sentence-Transformers library of the HuggingFace Model Hub. We also evaluated the accuracy of popular and recent LLMs, including their 4-bit quantizations, like falcon~\cite{refinedweb} and llama~\cite{touvron2023llama,touvron2023llama2}. Details of all tested models and their marks in this study are provided in \autoref{table:model&mark}. 

\noindent \textbf{Distance Metrics.} 
Our study primarily focuses on seven metrics: Cosine Distance, Euclidean Distance, Manhalanobis Distance~\cite{chandra1936generalised}, Braycurtis Distance, Lance and Williams Distance, Pearson Correlation Distance and Manhattan Distance. Cosine and Euclidean Distances are particularly prevalent in vector databases like Milvus~\cite{2021milvus}, FAISS~\cite{johnson2019billion}, Hnswlib~\cite{malkov2018efficient}, PGvector~\cite{pgvector}, Chroma~\cite{chroma}, DocArray~\cite{docarray}, Pinecone~\cite{pinecone}. All distance metrics are implemented by Scipy~\cite{scipy}.

\noindent \textbf{Baselines.} 
Text matching methods require simultaneous input of two sentences to assess semantic differences effectively, which means that the model needs to consider the semantics of both sentences together rather than independently. We've chosen three types of text matching methods to calculate text similarity as our baselines.
The first is QuestEval~\cite{scialom2021questeval}, an approach based on question generation and question answering (QG\&QA).
The second type involves cross-encoder models, like quora-distilroberta-base~\cite{Sanh2019DistilBERTAD}, nli-deberta-v3-base~\cite{he2021debertav3} and roberta-large-mnli~\cite{liu2019roberta}. The first model directly computes sentence similarity, while the latter two estimate probabilities of contradiction, neutral, and entailment. We take the probability of entailment as the similarity between texts. 
The third category includes extension methods based on NLI models, for which we selected SummaC~\cite{laban2022summac} as a baseline.

\noindent \textbf{Environment.} All experiments are performed on a workstation with Ubuntu 20.04.3 LTS and 4 NVIDIA V100 GPUs~(32GB memory).

\subsection{Evaluation metrics}

\textbf{Distance.} 
We measure sentence pair similarity using the distance between feature vectors. The distance from the base sentence to the positive sentence is noted as ``positive distance'', while its distance to the negative sentence is ``negative distance''. A smaller distance indicates higher similarity, with a distance of 0 signifying identical sentences. Ideally, the positive distance should be a number as small as possible, whereas the negative distance should be larger.

\noindent \textbf{Scores.} 
For those baselines, we interpret the scores generated by the methods as the similarity between two sentences. Scores associated with positive sentences are termed ``positive scores'', while those for negative sentences are ``negative scores''. Lower scores indicate greater similarity. We expect that positive scores will consistently be higher than negative scores.

\noindent \textbf{Accuracy.} In a test case, correct matching occurs if the method yields a smaller positive distance, or a higher positive score than the negative one. Incorrect matching happens in the opposite scenario. We evaluate a method's overall performance by tallying the number of correct matches across all test cases that should match.



\noindent \textbf{Average distance/score.} The representative distance/score for each dataset is determined by averaging the distances/scores between all sentence pairs in the set.

Note that all distances are calculated after normalizing the embedding vectors. All distances and scores in the experimental results are Max-Min normalized.

\subsection{RQ1: Are these test cases generated by our approach consistent with our requirements?}

\tool aim to construct complete triplets as test examples. Therefore, in this section, we evaluate whether the triplets satisfy the requirement of \autoref{eq:triplet}. We randomly sampled 100 cases from each dataset, creating 800 triplets with shuffled sentence orders. We labelled their semantic and structural similarities separately. We recruit three annotators with a Bachelor's degree or above and proficiency in both Chinese and English. The final labels are determined by majority vote. At first, annotators are asked to identify which two sentences had similar meanings. Finally, in 93.63\% of cases, the base and positive sentences are considered to be more semantically similar, aligning the semantic consistency requirement with \autoref{eq:triplet}.

After further shuffling the order within and between triplets, we posed a second question to them: Which two sentences do you think have a more similar sentence structure? We expected the base and negative sentences to be indexed as most structurally similar. With the exception of Err Nli, where negatives are derived from positives. The results align with our expectations in 92.75\% of the test cases, indicating that, in terms of sentence structure, the base sentences are typically more similar to the negative sentences than to the positive ones. 

A test case is deemed compliant if the responses to both questions align with our expectations. Our analysis shows that 86.5\% of the cases meet these criteria, suggesting that the majority of \tool triplets fulfill our construction objectives. The non-compliant fraction likely exhibits weaker semantic or structural properties, but not to the extent of complete incompatibility with our requirements. We excluded these cases from further experiments.

\addvspace{10pt}
\begin{mdframed}[backgroundcolor=gray!20]
    \textbf{RQ1 Answer:} The dataset constructed by \tool satisfies both structural and semantic requirements.
\end{mdframed}

\subsection{RQ2: Is \tool capable of identifying instances of false matching in vector databases}

\begin{table}[t]
\caption{The results of the case study}\vspace{-0.3cm}
\label{table:case-study}
\begin{adjustbox}{width=0.9\linewidth}
\begin{tabular}{ccc}
\toprule
Vector Database & Acc without transformation & Acc under MeTMaP \\ \midrule
\textbf{Annoy}           & 99.72                      & 37.72            \\
\textbf{Chroma}          & 70.28                      & 23.80            \\
\textbf{ScaNN}           & 99.72                      & 37.72            \\ \bottomrule
\end{tabular}
\end{adjustbox}
\end{table}

\begin{figure}[t]
\includegraphics[width=0.95\linewidth]{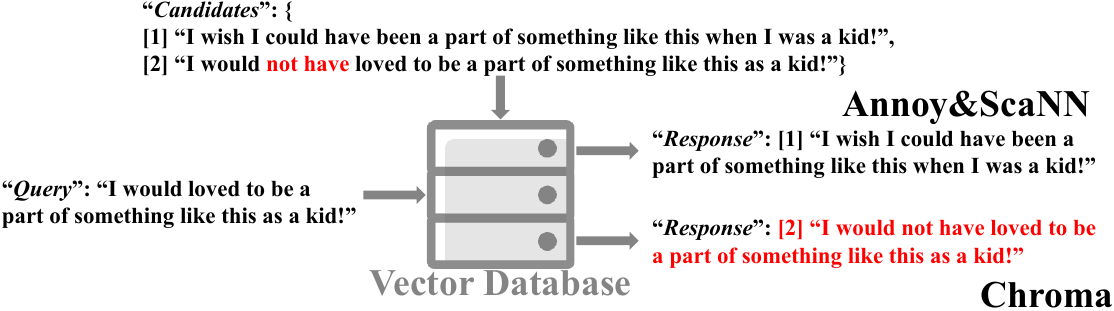}\vspace{-0.3cm}
\caption{An example of a real problem in a vector database.}
\label{fig:real-case}
\end{figure}

In this section, we conduct a case study to identify false matching problems in open source vector databases. The results of the study are shown in \autoref{table:case-study}. We observe that the query results of the vector databases demonstrate a high accuracy when no transformation are applied to the retrieved candidates. However, when candidates conform to our triplet specification, there is a noticeable drop in the accuracy of the database retrieval. An example of a real problem found in Chroma, is illustrated in \autoref{fig:real-case}. In this case, the negative sentence, formed by adding ``not have'' to the base sentence to convey opposite semantics, is incorrectly matched.  The response results from databases show that Annoy and ScaNN choose \textit{candidate-1} with the same semantics, whereas Chroma considers \textit{candidate-2} to be closer to the query, indicating a false matching problem.

Additionally, as Annoy and ScaNN utilize identical default embedding model and distance metric, they yield the same results. This suggest that components external to the matching method in these databases have minimal impact on retrieval outcomes.

\addvspace{10pt}
\begin{mdframed}[backgroundcolor=gray!20]
    \textbf{RQ2 Answer: \tool is effective in detecting a numerous false matching problems in real vector databases. The accuracy of these databases' retrieval largely depends on the vector matching methods employed, rather than on external integration components.}
\end{mdframed}

\subsection{RQ3: Can our approach find the erroneous outputs returned by the information retrieval methods?}


\begin{figure}[t]
    \centering
    \begin{subfigure}{0.419\linewidth}
        \centering
        \includegraphics[width=1\linewidth]{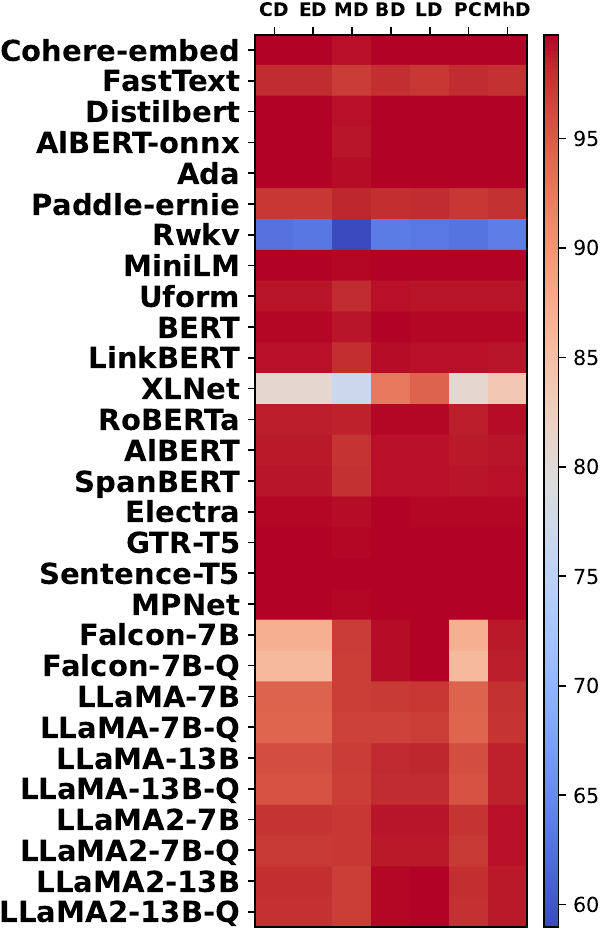}
        \caption{}
        \label{fig:distance1}
    \end{subfigure}
    \centering
    \begin{subfigure}{0.243\linewidth}
        \centering
        \includegraphics[width=1\linewidth]{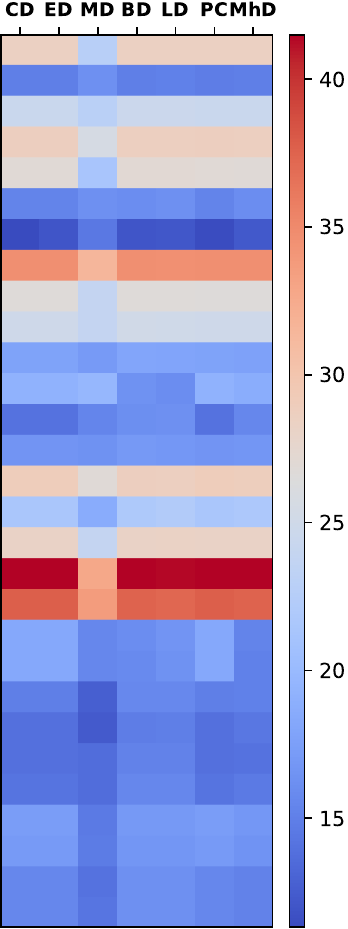}
        \caption{}
        \label{fig:distance2}
    \end{subfigure}
    \centering
    \begin{subfigure}{0.243\linewidth}
        \centering
        \includegraphics[width=1\linewidth]{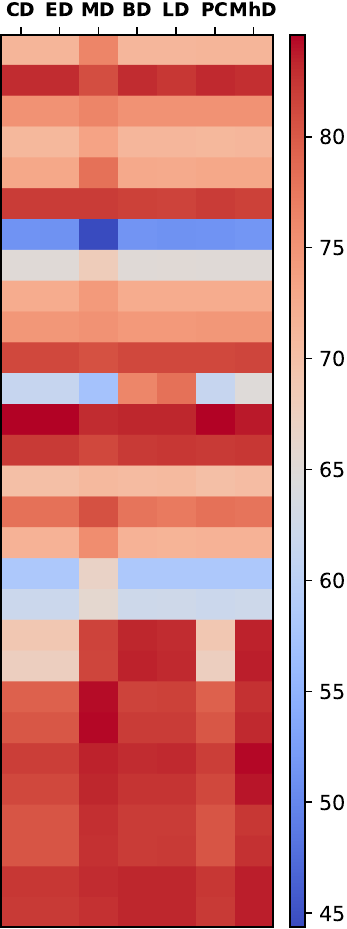}
        \caption{}
        \label{fig:distance3}
    \end{subfigure}
    \captionsetup{skip=10pt}\vspace{-0.3cm}
    \caption{The effect of metamorphosis on accuracy. \autoref{fig:distance1} shows the accuracy without transformation. \autoref{fig:distance2} gives the accuracy under \tool. \autoref{fig:distance3} is a statistic of the decrease in accuracy of the methods before and after the transformation.}\vspace{-0.3cm}
    \label{fig:distance-based acc}
\end{figure}

\tool is designed to construct triplet test cases to evaluate the accuracy of a specific information retrieval methods. 
We combine the embedding models and the distance metrics in pairs as various vector matching methods. Recognizing that developers might experiment with a range of combinations, even those not typically suited for practical vector matching, in this section, we evaluate 203 different vector matching methods. We quantify the matching performance of these methods on the entire dataset.

\autoref{fig:distance1} demonstrates that without additional metamorphosis, all methods exhibit high accuracy, with the lowest being 58.97\%. Remarkably, 177 methods achieve over 95\% accuracy, and 185 surpass 90\%. However, their performance drops significantly with the inclusion of disturbing sentences in test cases. As indicated in \autoref{fig:distance2} and \autoref{fig:distance3}, the average accuracy decreases by more than 78.07\%, with the highest accuracy among all combinations being only 41.51\%. Notably, 120 methods fall below 20\% accuracy. These findings suggest that \tool successfully uncovers numerous erroneous matching results produced by these methods.


Comparing with vector matching methods, these baselines demonstrate superior performance. \autoref{table:baseline total acc} shows that all six methods, along with their variants featuring reversed input order, outperform vector matching methods. Notably, eight of these achieve over 70\% accuracy even with interference, and five exceed 80\%. DeBERTa and its variant display an accuracy rate of 89\%, the best among all methods. These results further underscore \tool's effectiveness in detecting erroneous outputs in vector matching methods.
\begin{table}[t]
\caption{Accuracy of the Baselines on the total dataset}\vspace{-0.3cm}
\label{table:baseline total acc}
\begin{adjustbox}{width=0.5\linewidth}
\begin{tabular}{ccc}
\toprule
\textbf{Methods}       & \textbf{Default} & Reverse       \\ \midrule
\textbf{QuestEval}     & 0.56             & -    \\
\textbf{DistilRoBERTa} & 0.78             & -   \\
\textbf{DeBERTa}    & \textcolor{red}{0.89}    & \textcolor{red}{0.89} \\
\textbf{RoBERTa-mnli}  & 0.80             & 0.82 \\
\textbf{SummaC-zs}     & 0.79             & 0.82 \\
\textbf{SummaC-conv}   & 0.65             & 0.74 \\ \bottomrule
\end{tabular}
\end{adjustbox}\vspace{-0.3cm}
\end{table}

\addvspace{10pt}
\begin{mdframed}[backgroundcolor=gray!20]
    \textbf{RQ3 Answer: Under the test of \tool, all 203 vector matching methods have an accuracy of no more than 42\%, and 120 of them falling below 20\%. In contrast, the baselines performed better, with eight methods reaching 70\% or higher, and the best performer achieving 89\%.} 
\end{mdframed}

\subsection{RQ4: How different factors affect the performance of \tool?}

\begin{figure*}[t]
    \centering
    \begin{subfigure}{0.342\textwidth}
        \centering
        \includegraphics[width=1\textwidth]{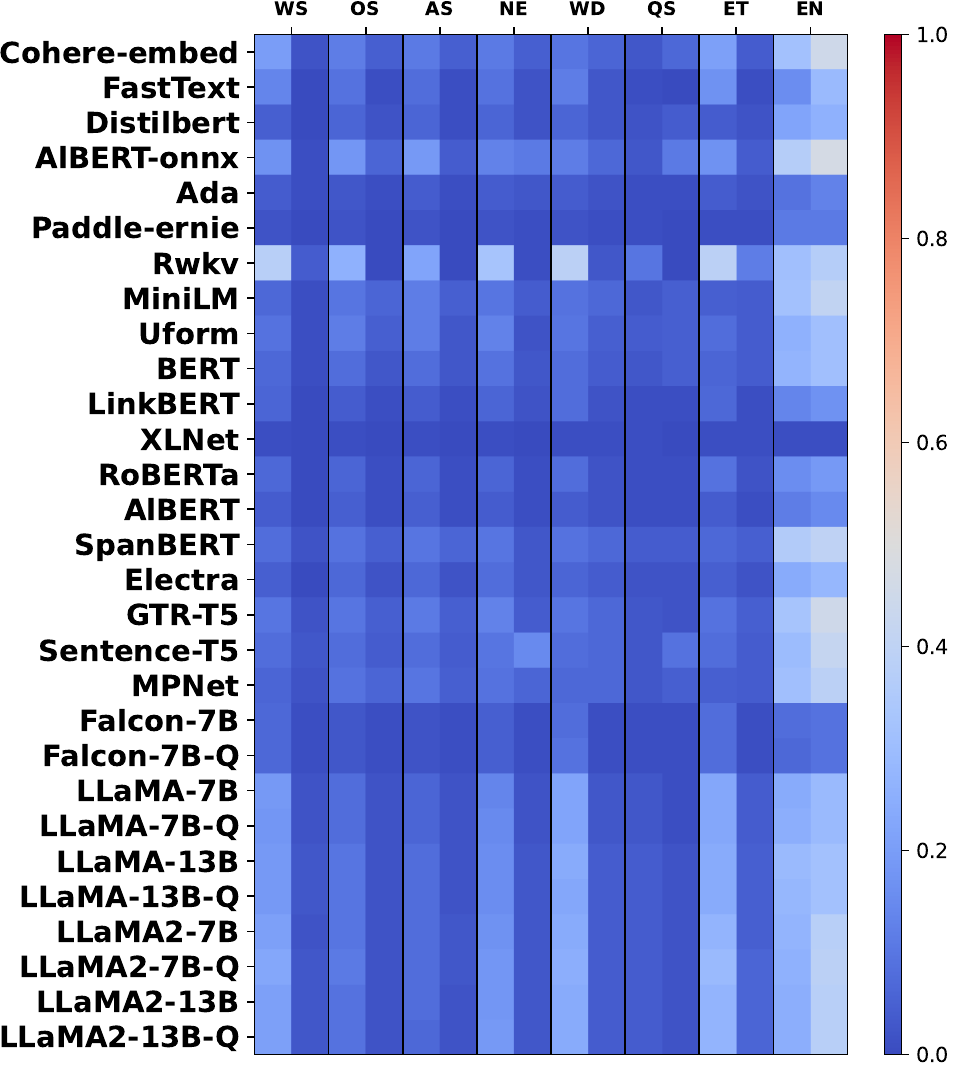}
        \caption{}
        \label{fig:mean-cd}
    \end{subfigure}
    \centering
    \begin{subfigure}{0.14\textwidth}
        \centering
        \includegraphics[width=1\textwidth]{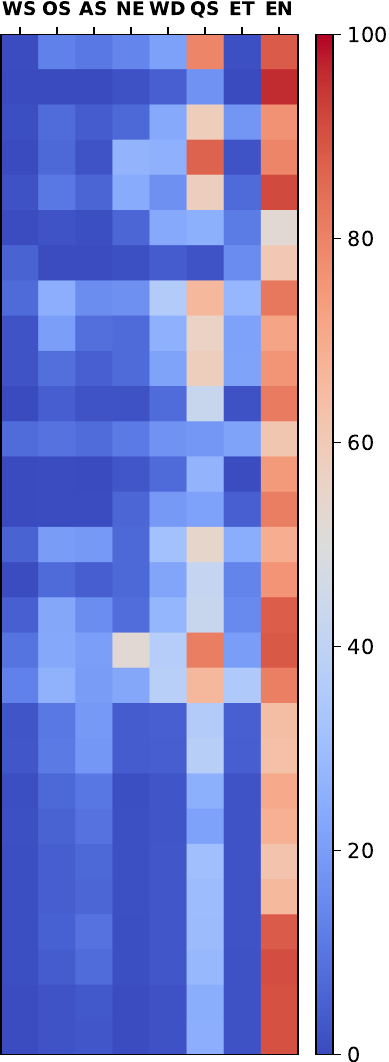}
        \caption{}
        \label{fig:acc-cd}
    \end{subfigure}
    \begin{subfigure}{0.252\textwidth}
        \centering
        \includegraphics[width=1\textwidth]{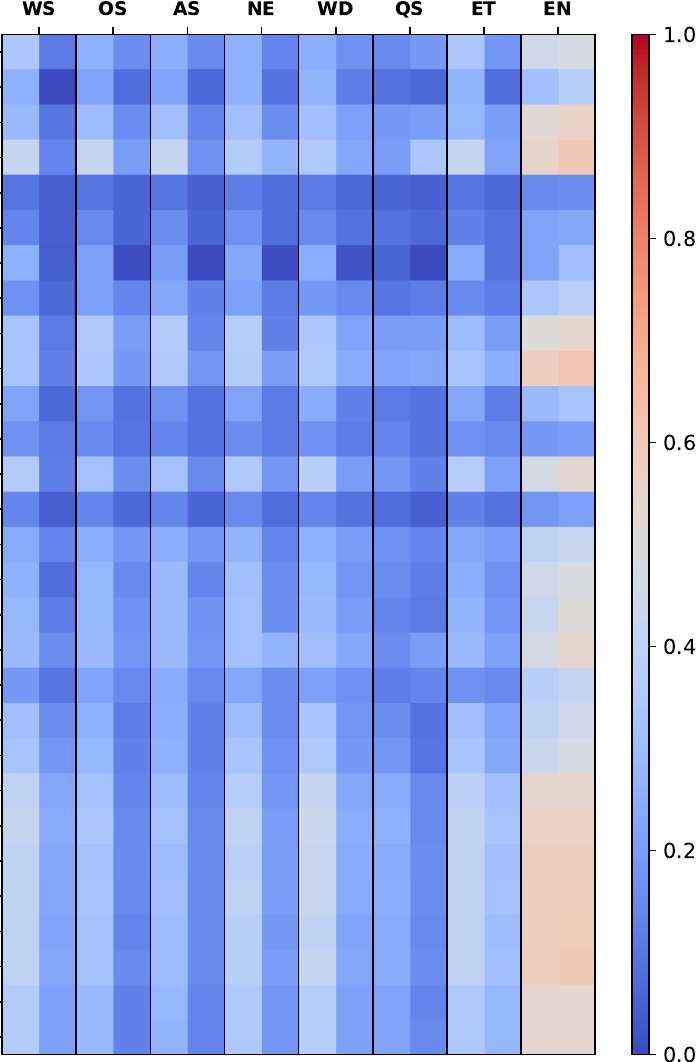}
        \caption{}
        \label{fig:mean-md}
    \end{subfigure}
    \begin{subfigure}{0.14\textwidth}
        \centering
        \includegraphics[width=1\textwidth]{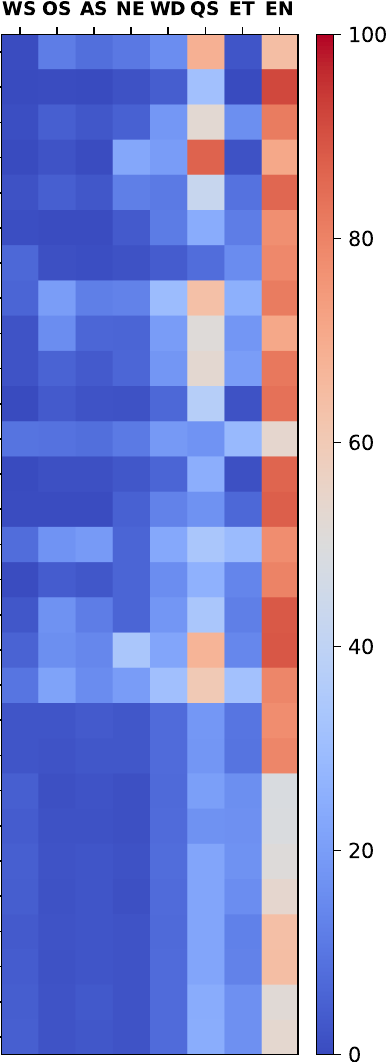}
        \caption{}
        \label{fig:acc-md}
    \end{subfigure}
    \captionsetup{skip=10pt}\vspace{-0.2cm}
    \caption{Average positive and negative distances and accuracies on all sub-datasets.\autoref{fig:mean-cd} shows the average positive and negative distances for the CD-based methods, where the left side of each column representes the positive one, and the right is the negative one. \autoref{fig:acc-cd} is the accuracy of all CD-based methods on sub-datasets. \autoref{fig:mean-md} shows the average distances for the MD-based methods. \autoref{fig:acc-md} is the accuracy of all MD-based methods on sub-datasets.} \vspace{-0.2cm}
    \label{fig:cd&md-based result}
\end{figure*}



As shown in \autoref{fig:cd&md-based result}, when \tool employs different MRs for metamorphosis, the vector matching methods produce significant variation in the number of erroneous outputs. Specifically, the distribution of average negative distances shows remarkable similarity across the seven distance metrics. Notably, the lowest values frequently appear in Word Swap, Obj Sub, Quant Sub, and Act Sub categories. For instance, the percentage of minima occurring in these four categories is 72.41\%, 41.38\%, 55.17\%, 34.48\% in CD and 55.17\%, 27.59\%, 31.03\% and 27.59\% in MD, respectively. 
Furthermore, in word-level MRs, positive mean distances are typically exceed negative mean distances. 

Additionally, \autoref{fig:cd&md-based result} also indicates a strong correlation between the lowest matching accuracy of each model and specific categories. Specifically, lower average negative distances often correspond to lower accuracy. We consider that this trend relates to the degree of structural change induced by metamorphosis. In Word Swap, where only two words are swapped without altering sentence structure or introducing new words, the metamorphosis is minimal. In contrast, MRs based on word substitution, while not altering structure, introduce new words, resulting in more significant metamorphosis compared to Word Swap. The most substantial structural changes at the word level occur in Nega Exp and Word Del, as these involve adding negatives or deleting parts of sentences.
At the sentence level, Back Translation usually involves a combination of multiple word level MRs, resulting in more extensive structural changes than those at the word level. In Inference Generation, base and positive sentences differ significantly in structure as they are connected by inference. These difference between base and negative sentence is further amplified by removing evidence supporting the base sentence from the positive sentence. We believe that is the highest negative average distances and peak matching accuracies consistently occur in this category across all distance metrics.

Overall, we consider that vector matching methods are more sensitive to structural variations in sentences than to semantic differences. \tool detects a higher number of erroneous outputs when evaluating pairs with less structural metamorphosis.

As discussed in \autoref{sec:methodology}, Word Deletion and Inference Generation exhibit \textit{One-Way Inconsistency} between base sentences and their respective negative and positive sentences. These inconsistencies are detectable only in a specific direction. Vector matching methods, however, tend to overlook these nuances due to their disregard for input order. We expect these methods, which have requirements on the order of inputs, to detect these problems. For methods sensitive to input sequence, we frame their task as determining the likelihood of \textit{sentence1} entailing \textit{sentence2}, assuming \textit{sentence1} is the first input.

\begin{figure}[t]
    \centering
    \begin{subfigure}{0.684\linewidth}
        \centering
        \includegraphics[width=1\linewidth]{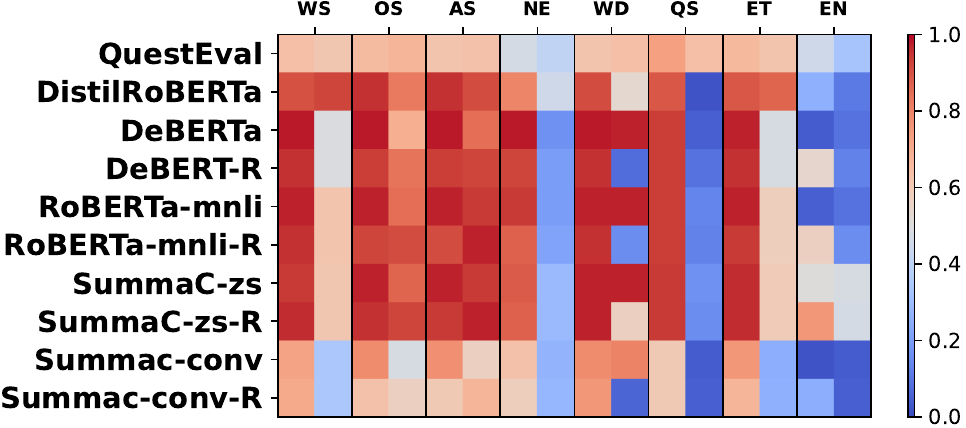}
        \caption{}
        \label{fig:mean-score}
    \end{subfigure}
    \centering
    \begin{subfigure}{0.271\linewidth}
        \centering
        \includegraphics[width=1\linewidth]{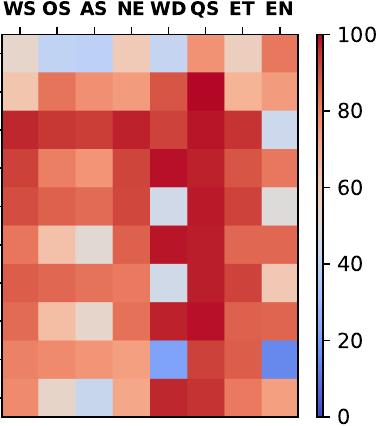}
        \caption{}
        \label{fig:acc-score}
    \end{subfigure}
    \captionsetup{skip=10pt}\vspace{-0.3cm}
    \caption{Average positive and negative scores and accuracies of Baselines on all sub-datasets. \autoref{fig:mean-score} shows the average positive and negative scores, where the left side of each column represents the positive one, and the right is the negative one. \autoref{fig:acc-score} is the accuracy of baselines.}\vspace{-0.1cm}
    \label{fig:baseline result}
\end{figure}

As we see in \autoref{fig:baseline result}, NLI model-based methods and cross-encoder models typically yield higher negative scores for Word Del and lower positive scores for Err Nli, with the exception of DistilRoBERTa. This is due to the fact that DistilRoBERTa differs as it evaluates the similarity of two sentences regardless of their input order. Variants using a consistent-with-expectations approach produce contrasting results, indicating their capability to detect one-way inconsistency. This also highlights that NLI-based text matching methods are more adept at uncovering semantic differences. Although inconsistencies can be detected, it is the positive sentence that should be closer to the base one in Word Del cases, as it encompass a comprehensive semantics of the base sentence.

\addvspace{10pt}
\begin{mdframed}[backgroundcolor=gray!20]
    \textbf{RQ4 Answer: \tool identifies more erroneous outputs when testing with cases constructed via minor metamorphosis. Additionally, by specifying the input order of sentences, \tool uncovers new erroneous outputs by revealing one-way inconsistencies in specific scenarios.} 
\end{mdframed}
\section{Discussions}
\label{sec:discuss}

\noindent\textbf{Limitations on proposed metamorphic relations.} 
 A limitation may arise from the use of MRs derived from prominent datasets, which may not fully encompass all word and sentence-level metamorphosis. The metamorphosis we chose intentionally minimally alter sentence structure to gauge their impact on information retrieval accuracy. This focus on subtle shifts aims to reveal how minor variations can affect retrieval robustness and efficiency, despite limiting the range of metamorphosis types considered.

\noindent\textbf{Sensitivity to MRs.} 
Our study might also be limited by the varying sensitivities of models to different MRs. Some models may be better at detecting certain metamorphosis, like word swaps, than others, such as quantifier changes.  Nonetheless, our study's primary focus was on the rules themselves. The differential sensitivity to these rules, while an important aspect, is considered orthogonal to our primary scope.


\noindent\textbf{Potential for Triplet Generation.} 
Currently, only the triplet corresponding to Quant Sub is generated from a base sentence. However, there is significant potential to extend this functionality to include triplet generation for other categories. Our goal is to develop a system capable of automatically generating multiple triplets for various MRs from any given sentence, thus greatly enhancing the robustness and versatility of our approach.

\noindent\textbf{Text-matching versus Vector-matching.} 
Our research highlights a key limitation involving the trade-off between text-matching and vector matching methods. Text-matching offers higher accuracy but is time-intensive, requiring the calculation of similarity between the query and every sentence in the database for each retrieval. Vector matching, on the other hand, is faster and simpler to implement but sometimes less accurate. We propose a future solution that combines both: using vector matching first to quickly identify a subset of potential matches, followed by text-matching for precise final match identification. This integrated approach could potentially capitalize on the strengths of both methods, achieving a balance of speed and accuracy.

\noindent\textbf{False Matching Mitigation.} Following our discussions with vector database developers regarding the false matching issue in vector databases, we're considering a combined approach that leverages the strengths of both vector and text matching methods. This method would initially group candidates using the quick vector matching process. Then, we'd select the most similar group and apply the more accurate text matching method to score and identify the best answer within this group. However, it's important to note that this solution requires the databases to support both vector and text matching methods for retrieval.

\vspace{0.2cm}
\section{Threats To Validity}
\noindent\textbf{Internal Threat.}
In assessing our study's validity, the main internal threats stem from our dataset's reliability. Although we took stringent steps to ensure accuracy, including having three co-authors meticulously label each sentence's semantics in our dataset, there's still a possibility of bias and inaccuracies. Even minor oversights or unconscious biases could impact our analyses, influencing our results and interpretations. Acknowledging this limitation, future efforts should include a broader range of labeling perspectives or adopt automated validation methods to further reduce these risks.

\noindent\textbf{External Threat.}
When assessing external threats to validity, a significant concern arises from the model quantization. To accommodate models with a larger number of parameters on limited video memory, we employed a 4-bit quantization technique. However, this approach inherently leads to a loss of precision, which can adversely affect the method's accuracy. The quantization process, while efficient in terms of memory usage, might introduce errors or deviations due to the reduced bit representation, potentially limiting the generalizability or accuracy of our findings when applied to models that aren't subjected to similar quantization.

\vspace{0.2cm}
\section{Related Work}
\label{sec:relwork}
\subsection{Testing NLP Applications}
Natural Language Processing~(NLP) applications are extensively used in tasks such as information retrieval~\cite{lewis1996natural,voorhees1999natural,yew2023transforming}, sentiment analysis ~\cite{william2023natural,sait2023deep,fanni2023natural}, virtual assistants~\cite{hariri2023unlocking,zhou2023artificially,kamoonpuri2023hi}, and text/code summarization~\cite{luo2023chatgpt,yang2023exploring,geng2023interpretation}. Given this diversity, many studies have focused on assessing the robustness and reliability of NLP applications. In contrast, our work specifically targets two aspects of LLM-augmented generation applications, areas not yet systematically evaluated in existing literature.


\noindent\textbf{Robustness.}
Ribeiro et al.~\cite{ribeiro-etal-2020-beyond} developed a behavioral testing method for NLP, focusing on sentiment analysis, duplicate question answering, and comprehension tasks.
Li et al.~\cite{LiJDLW19} utilized deep learning to generate test cases for deep learning-driven NLP applications.
Sun et al.~\cite{Sun2022ImprovingMT} introduced a word replacement technique for correcting machine translation errors without retraining.
Xiao et al.~\cite{xiao2023leap} proposed an automated testing approach combining LEvy flight-based Adaptive Particle Swarm Optimization with text analysis for creating adversarial scenarios.
Wang et al.~\cite{Wang_ICSE23} conducted a comprehensive study on the robustness of popular content moderation software.


\noindent\textbf{Reliability.}
Tan et al.~\cite{tan2021reliability} propose a framework for realiability testing in NLP systems, using adversarial attacks and interdisciplinary collaboration to uphold industry standards. 
Chen et al.~\cite{Chen_ASE2021} introduce a label-independent testing method for QA software through Metamorphic Relations applied to recursive questions, uncovering prevalent answering flaws.
Alshemali et al.~\cite{alshemali2020improving} reviews the vulnerability of deep neural networks in NLP to adversarial examples, categorizes methods for generating adversarial texts, and explores defensive strategies and challenges.

\vspace{0.2cm}
\subsection{Testing LLM-integrated Applications}
\cite{hou2023large} explored the intersection of LLMs and software engineering.
Testing LLM-integrated applications is fraught with challenges and complexities. The rapid adoption of OpenAI's GPT API highlighted these issues for both developers and integrators. Libraries such as Langchain~\cite{langchain}, ChatDB~\cite{hu2023chatdb}, LlamaIndex~\cite{Liu_LlamaIndex_2022}, and Flowise~\cite{Flowise} simplify the development and testing of LLM-integrated applications by converting language into more testable models.

Prior research primarily tests LLM-integrated applications for potential attack vectors. Techniques have been crafted to circumvent application safeguards, known as ``jailbreaking'' ~\cite{Dan,deng2023jailbreaker}, hijack the prompt's intent~\cite{liu2023prompt}, or even leak the prompt itself~\cite{perez2022ignore}. These applications also undergo testing for vulnerability to backdoor attacks, where tainted datasets trigger specific responses or malicious behaviors~\cite{guo2022threats,li2021hidden}. A novel threat, ``indirect prompt injections'', allows applications to unwittingly execute commands from external API calls, such as SQL query results~\cite{greshake2023youve}. 
Furthermore, \cite{li2024digger} investigated the copyright issues of data used in the training process of LLMs. Complementing these studies, our work focuses on the interaction between LLMs and their integrated memory, examining methods for accurate information matching and retrieval.


\vspace{0.3cm}
\section{Conclusion}
\label{sec:conclusion}

This paper introduces a groundbreaking metamorphic testing framework tailored for LLM augmented generation, pioneering the evaluation of vector matching techniques in this context. The comprehensive assessment of 203 vector matching methods reveals a notable bias towards structural similarity, underscoring a critical area for improvement in semantic accuracy. These findings not only highlight the existing challenges in LLM augmentation but also provide a clear direction for future enhancements. As we continue to refine these techniques, our work lays a foundational path towards developing more accurate, reliable, and semantically sophisticated LLM augmented systems.

\clearpage
\bibliographystyle{ACM-Reference-Format}
\bibliography{bib}


\begin{thebibliography}{00}


\ifx \showCODEN    \undefined \def \showCODEN     #1{\unskip}     \fi
\ifx \showDOI      \undefined \def \showDOI       #1{#1}\fi
\ifx \showISBNx    \undefined \def \showISBNx     #1{\unskip}     \fi
\ifx \showISBNxiii \undefined \def \showISBNxiii  #1{\unskip}     \fi
\ifx \showISSN     \undefined \def \showISSN      #1{\unskip}     \fi
\ifx \showLCCN     \undefined \def \showLCCN      #1{\unskip}     \fi
\ifx \shownote     \undefined \def \shownote      #1{#1}          \fi
\ifx \showarticletitle \undefined \def \showarticletitle #1{#1}   \fi
\ifx \showURL      \undefined \def \showURL       {\relax}        \fi
\providecommand\bibfield[2]{#2}
\providecommand\bibinfo[2]{#2}
\providecommand\natexlab[1]{#1}
\providecommand\showeprint[2][]{arXiv:#2}

\bibitem[\protect\citeauthoryear{??}{our}{2023}]%
        {ourtool}
 \bibinfo{year}{2023}\natexlab{}.
\newblock \bibinfo{title}{\tool{}}.
\newblock \bibinfo{howpublished}{\url{https://anonymous.4open.science/r/MeTMaP-879B}}.   (\bibinfo{year}{2023}).
\newblock


\bibitem[\protect\citeauthoryear{0xk1h0}{0xk1h0}{2023}]%
        {Dan}
\bibfield{author}{\bibinfo{person}{0xk1h0}.} \bibinfo{year}{2023}\natexlab{}.
\newblock \bibinfo{title}{{ChatGPT\_DAN}}.
\newblock \bibinfo{howpublished}{\url{https://github.com/0xk1h0/ChatGPT_DAN}}.   (\bibinfo{year}{2023}).
\newblock


\bibitem[\protect\citeauthoryear{Alshemali and Kalita}{Alshemali and Kalita}{2020}]%
        {alshemali2020improving}
\bibfield{author}{\bibinfo{person}{Basemah Alshemali} {and} \bibinfo{person}{Jugal Kalita}.} \bibinfo{year}{2020}\natexlab{}.
\newblock \showarticletitle{{Improving the reliability of deep neural networks in NLP: A review}}.
\newblock \bibinfo{journal}{{\em {Knowledge-Based Systems}\/}}  \bibinfo{volume}{191} (\bibinfo{year}{2020}).
\newblock


\bibitem[\protect\citeauthoryear{Chandra et~al\mbox{.}}{Chandra et~al\mbox{.}}{1936}]%
        {chandra1936generalised}
\bibfield{author}{\bibinfo{person}{Mahalanobis~Prasanta Chandra} {et~al\mbox{.}}} \bibinfo{year}{1936}\natexlab{}.
\newblock \showarticletitle{{On the generalised distance in statistics}}. In \bibinfo{booktitle}{{\em {Proceedings of the National Institute of Sciences of India}}}, Vol.~\bibinfo{volume}{2}. \bibinfo{pages}{49--55}.
\newblock


\bibitem[\protect\citeauthoryear{Chase}{Chase}{2022}]%
        {langchain}
\bibfield{author}{\bibinfo{person}{Harrison Chase}.} \bibinfo{year}{2022}\natexlab{}.
\newblock \bibinfo{title}{{LangChain}}.
\newblock \bibinfo{howpublished}{\url{https://python.langchain.com/docs/get_started/introduction}}.   (\bibinfo{year}{2022}).
\newblock


\bibitem[\protect\citeauthoryear{Chen, Jin, and Xie}{Chen et~al\mbox{.}}{2021}]%
        {Chen_ASE2021}
\bibfield{author}{\bibinfo{person}{Songqiang Chen}, \bibinfo{person}{Shuo Jin}, {and} \bibinfo{person}{Xiaoyuan Xie}.} \bibinfo{year}{2021}\natexlab{}.
\newblock \showarticletitle{{Testing Your Question Answering Software via Asking Recursively}}. In \bibinfo{booktitle}{{\em {ASE}}}. \bibinfo{pages}{104--116}.
\newblock
\showDOI{%
\url{https://doi.org/10.1109/ASE51524.2021.9678670}}


\bibitem[\protect\citeauthoryear{Chen, Cheung, and Yiu}{Chen et~al\mbox{.}}{2020}]%
        {Chen2020MetamorphicTA}
\bibfield{author}{\bibinfo{person}{Tsong~Yueh Chen}, \bibinfo{person}{S.~C. Cheung}, {and} \bibinfo{person}{Siu-Ming Yiu}.} \bibinfo{year}{2020}\natexlab{}.
\newblock \showarticletitle{{Metamorphic Testing: A New Approach for Generating Next Test Cases}}.
\newblock \bibinfo{journal}{{\em {ArXiv}\/}}  \bibinfo{volume}{abs/2002.12543} (\bibinfo{year}{2020}).
\newblock
\showURL{%
\url{https://api.semanticscholar.org/CorpusID:15467386}}


\bibitem[\protect\citeauthoryear{Chen, Kuo, Liu, Poon, Towey, Tse, and Zhou}{Chen et~al\mbox{.}}{2018}]%
        {ChenTY2018}
\bibfield{author}{\bibinfo{person}{Tsong~Yueh Chen}, \bibinfo{person}{Fei-Ching Kuo}, \bibinfo{person}{Huai Liu}, \bibinfo{person}{Pak-Lok Poon}, \bibinfo{person}{Dave Towey}, \bibinfo{person}{T.~H. Tse}, {and} \bibinfo{person}{Zhi~Quan Zhou}.} \bibinfo{year}{2018}\natexlab{}.
\newblock \showarticletitle{{Metamorphic Testing: A Review of Challenges and Opportunities}}.
\newblock \bibinfo{journal}{{\em {ACM Comput. Surv.}\/}} \bibinfo{volume}{51}, \bibinfo{number}{1} (\bibinfo{date}{jan} \bibinfo{year}{2018}), 27.
\newblock


\bibitem[\protect\citeauthoryear{Chiang, Chuang, Glass, and Lee}{Chiang et~al\mbox{.}}{2023}]%
        {chiang2023revealing}
\bibfield{author}{\bibinfo{person}{Cheng-Han Chiang}, \bibinfo{person}{Yung-Sung Chuang}, \bibinfo{person}{James Glass}, {and} \bibinfo{person}{Hung-yi Lee}.} \bibinfo{year}{2023}\natexlab{}.
\newblock \showarticletitle{{Revealing the Blind Spot of Sentence Encoder Evaluation by HEROS}}.
\newblock \bibinfo{journal}{{\em {arXiv preprint arXiv:2306.05083}\/}} (\bibinfo{year}{2023}).
\newblock


\bibitem[\protect\citeauthoryear{chroma core}{chroma core}{2023}]%
        {chroma}
\bibfield{author}{\bibinfo{person}{chroma core}.} \bibinfo{year}{2023}\natexlab{}.
\newblock \bibinfo{title}{{Chroma}}.
\newblock \bibinfo{howpublished}{\url{https://github.com/chroma-core/chroma}}.   (\bibinfo{year}{2023}).
\newblock


\bibitem[\protect\citeauthoryear{Clark, Luong, Le, and Manning}{Clark et~al\mbox{.}}{2020}]%
        {clark2020electra}
\bibfield{author}{\bibinfo{person}{Kevin Clark}, \bibinfo{person}{Minh-Thang Luong}, \bibinfo{person}{Quoc~V Le}, {and} \bibinfo{person}{Christopher~D Manning}.} \bibinfo{year}{2020}\natexlab{}.
\newblock \showarticletitle{{Electra: Pre-training text encoders as discriminators rather than generators}}.
\newblock \bibinfo{journal}{{\em {arXiv preprint arXiv:2003.10555}\/}} (\bibinfo{year}{2020}).
\newblock


\bibitem[\protect\citeauthoryear{cohere}{cohere}{2023}]%
        {cohere}
\bibfield{author}{\bibinfo{person}{cohere}.} \bibinfo{year}{2023}\natexlab{}.
\newblock \bibinfo{title}{{Cohere}}.
\newblock \bibinfo{howpublished}{\url{https://dashboard.cohere.com/}}.   (\bibinfo{year}{2023}).
\newblock


\bibitem[\protect\citeauthoryear{De~Marneffe, Rafferty, and Manning}{De~Marneffe et~al\mbox{.}}{2008}]%
        {de2008finding}
\bibfield{author}{\bibinfo{person}{Marie-Catherine De~Marneffe}, \bibinfo{person}{Anna~N Rafferty}, {and} \bibinfo{person}{Christopher~D Manning}.} \bibinfo{year}{2008}\natexlab{}.
\newblock \showarticletitle{{Finding contradictions in text}}. In \bibinfo{booktitle}{{\em {Proceedings of acl-08: Hlt}}}. \bibinfo{pages}{1039--1047}.
\newblock


\bibitem[\protect\citeauthoryear{Deng, Liu, Li, Wang, Zhang, Li, Wang, Zhang, and Liu}{Deng et~al\mbox{.}}{2023}]%
        {deng2023jailbreaker}
\bibfield{author}{\bibinfo{person}{Gelei Deng}, \bibinfo{person}{Yi Liu}, \bibinfo{person}{Yuekang Li}, \bibinfo{person}{Kailong Wang}, \bibinfo{person}{Ying Zhang}, \bibinfo{person}{Zefeng Li}, \bibinfo{person}{Haoyu Wang}, \bibinfo{person}{Tianwei Zhang}, {and} \bibinfo{person}{Yang Liu}.} \bibinfo{year}{2023}\natexlab{}.
\newblock \bibinfo{title}{{Jailbreaker: Automated Jailbreak Across Multiple Large Language Model Chatbots}}.
\newblock   (\bibinfo{year}{2023}).
\newblock
\showeprint[{arXiv}]{cs.CR/2307.08715}


\bibitem[\protect\citeauthoryear{Devlin, Chang, Lee, and Toutanova}{Devlin et~al\mbox{.}}{2018}]%
        {DBLP:journals/corr/abs-1810-04805}
\bibfield{author}{\bibinfo{person}{Jacob Devlin}, \bibinfo{person}{Ming{-}Wei Chang}, \bibinfo{person}{Kenton Lee}, {and} \bibinfo{person}{Kristina Toutanova}.} \bibinfo{year}{2018}\natexlab{}.
\newblock \showarticletitle{{{BERT:} Pre-training of Deep Bidirectional Transformers for Language Understanding}}.
\newblock \bibinfo{journal}{{\em {CoRR}\/}}  \bibinfo{volume}{abs/1810.04805} (\bibinfo{year}{2018}).
\newblock
\showeprint[arxiv]{1810.04805}
\showURL{%
\url{http://arxiv.org/abs/1810.04805}}


\bibitem[\protect\citeauthoryear{docarray}{docarray}{2023}]%
        {docarray}
\bibfield{author}{\bibinfo{person}{docarray}.} \bibinfo{year}{2023}\natexlab{}.
\newblock \bibinfo{title}{{DocArray}}.
\newblock \bibinfo{howpublished}{\url{https://github.com/docarray/docarray}}.   (\bibinfo{year}{2023}).
\newblock


\bibitem[\protect\citeauthoryear{Fanni, Febi, Aghakhanyan, and Neri}{Fanni et~al\mbox{.}}{2023}]%
        {fanni2023natural}
\bibfield{author}{\bibinfo{person}{Salvatore~Claudio Fanni}, \bibinfo{person}{Maria Febi}, \bibinfo{person}{Gayane Aghakhanyan}, {and} \bibinfo{person}{Emanuele Neri}.} \bibinfo{year}{2023}\natexlab{}.
\newblock \showarticletitle{{Natural language processing}}.
\newblock In \bibinfo{booktitle}{{\em {Introduction to Artificial Intelligence}}}. \bibinfo{pages}{87--99}.
\newblock


\bibitem[\protect\citeauthoryear{FlowiseAI}{FlowiseAI}{2023}]%
        {Flowise}
\bibfield{author}{\bibinfo{person}{FlowiseAI}.} \bibinfo{year}{2023}\natexlab{}.
\newblock \bibinfo{title}{{Flowise}}.
\newblock \bibinfo{howpublished}{\url{https://github.com/FlowiseAI/Flowise}}.   (\bibinfo{year}{2023}).
\newblock


\bibitem[\protect\citeauthoryear{Geng, Wang, Dong, Wang, Cao, Zhang, and Jin}{Geng et~al\mbox{.}}{2023}]%
        {geng2023interpretation}
\bibfield{author}{\bibinfo{person}{Mingyang Geng}, \bibinfo{person}{Shangwen Wang}, \bibinfo{person}{Dezun Dong}, \bibinfo{person}{Haotian Wang}, \bibinfo{person}{Shaomeng Cao}, \bibinfo{person}{Kechi Zhang}, {and} \bibinfo{person}{Zhi Jin}.} \bibinfo{year}{2023}\natexlab{}.
\newblock \showarticletitle{{Interpretation-based Code Summarization}}. In \bibinfo{booktitle}{{\em ICPC}}.
\newblock


\bibitem[\protect\citeauthoryear{Greshake, Abdelnabi, Mishra, Endres, Holz, and Fritz}{Greshake et~al\mbox{.}}{2023}]%
        {greshake2023youve}
\bibfield{author}{\bibinfo{person}{Kai Greshake}, \bibinfo{person}{Sahar Abdelnabi}, \bibinfo{person}{Shailesh Mishra}, \bibinfo{person}{Christoph Endres}, \bibinfo{person}{Thorsten Holz}, {and} \bibinfo{person}{Mario Fritz}.} \bibinfo{year}{2023}\natexlab{}.
\newblock \bibinfo{title}{{Not what you've signed up for: Compromising Real-World LLM-Integrated Applications with Indirect Prompt Injection}}.
\newblock   (\bibinfo{year}{2023}).
\newblock
\showeprint[{arXiv}]{cs.CR/2302.12173}


\bibitem[\protect\citeauthoryear{Gugger}{Gugger}{2023}]%
        {rwkv}
\bibfield{author}{\bibinfo{person}{Sylvain Gugger}.} \bibinfo{year}{2023}\natexlab{}.
\newblock \bibinfo{title}{{RWKV}}.
\newblock \bibinfo{howpublished}{\url{https://huggingface.co/sgugger/rwkv-430M-pile}}.   (\bibinfo{year}{2023}).
\newblock


\bibitem[\protect\citeauthoryear{Guo, Sun, Lindgren, Geng, Simcha, Chern, and Kumar}{Guo et~al\mbox{.}}{2020}]%
        {scann}
\bibfield{author}{\bibinfo{person}{Ruiqi Guo}, \bibinfo{person}{Philip Sun}, \bibinfo{person}{Erik Lindgren}, \bibinfo{person}{Quan Geng}, \bibinfo{person}{David Simcha}, \bibinfo{person}{Felix Chern}, {and} \bibinfo{person}{Sanjiv Kumar}.} \bibinfo{year}{2020}\natexlab{}.
\newblock \showarticletitle{{Accelerating Large-Scale Inference with Anisotropic Vector Quantization}}. In \bibinfo{booktitle}{{\em International Conference on Machine Learning}}.
\newblock
\showURL{%
\url{https://arxiv.org/abs/1908.10396}}


\bibitem[\protect\citeauthoryear{Guo, Xie, Li, Lyu, and Zhang}{Guo et~al\mbox{.}}{2022}]%
        {guo2022threats}
\bibfield{author}{\bibinfo{person}{Shangwei Guo}, \bibinfo{person}{Chunlong Xie}, \bibinfo{person}{Jiwei Li}, \bibinfo{person}{Lingjuan Lyu}, {and} \bibinfo{person}{Tianwei Zhang}.} \bibinfo{year}{2022}\natexlab{}.
\newblock \bibinfo{title}{{Threats to Pre-trained Language Models: Survey and Taxonomy}}.
\newblock   (\bibinfo{year}{2022}).
\newblock
\showeprint[{arXiv}]{cs.CR/2202.06862}


\bibitem[\protect\citeauthoryear{Hariri}{Hariri}{2023}]%
        {hariri2023unlocking}
\bibfield{author}{\bibinfo{person}{Walid Hariri}.} \bibinfo{year}{2023}\natexlab{}.
\newblock \showarticletitle{{Unlocking the Potential of ChatGPT: A Comprehensive Exploration of its Applications, Advantages, Limitations, and Future Directions in Natural Language Processing}}.
\newblock \bibinfo{journal}{{\em {arXiv preprint}\/}} (\bibinfo{year}{2023}).
\newblock


\bibitem[\protect\citeauthoryear{He, Gao, and Chen}{He et~al\mbox{.}}{2021}]%
        {he2021debertav3}
\bibfield{author}{\bibinfo{person}{Pengcheng He}, \bibinfo{person}{Jianfeng Gao}, {and} \bibinfo{person}{Weizhu Chen}.} \bibinfo{year}{2021}\natexlab{}.
\newblock \bibinfo{title}{{DeBERTaV3: Improving DeBERTa using ELECTRA-Style Pre-Training with Gradient-Disentangled Embedding Sharing}}.
\newblock   (\bibinfo{year}{2021}).
\newblock
\showeprint[{arXiv}]{cs.CL/2111.09543}


\bibitem[\protect\citeauthoryear{Hou, Zhao, Liu, Yang, Wang, Li, Luo, Lo, Grundy, and Wang}{Hou et~al\mbox{.}}{2023}]%
        {hou2023large}
\bibfield{author}{\bibinfo{person}{Xinyi Hou}, \bibinfo{person}{Yanjie Zhao}, \bibinfo{person}{Yue Liu}, \bibinfo{person}{Zhou Yang}, \bibinfo{person}{Kailong Wang}, \bibinfo{person}{Li Li}, \bibinfo{person}{Xiapu Luo}, \bibinfo{person}{David Lo}, \bibinfo{person}{John Grundy}, {and} \bibinfo{person}{Haoyu Wang}.} \bibinfo{year}{2023}\natexlab{}.
\newblock \bibinfo{title}{{Large Language Models for Software Engineering: A Systematic Literature Review}}.
\newblock   (\bibinfo{year}{2023}).
\newblock
\showeprint[arxiv]{cs.SE/2308.10620}


\bibitem[\protect\citeauthoryear{Hu, Fu, Du, Luo, Zhao, and Zhao}{Hu et~al\mbox{.}}{2023}]%
        {hu2023chatdb}
\bibfield{author}{\bibinfo{person}{Chenxu Hu}, \bibinfo{person}{Jie Fu}, \bibinfo{person}{Chenzhuang Du}, \bibinfo{person}{Simian Luo}, \bibinfo{person}{Junbo Zhao}, {and} \bibinfo{person}{Hang Zhao}.} \bibinfo{year}{2023}\natexlab{}.
\newblock \bibinfo{title}{{ChatDB: Augmenting LLMs with Databases as Their Symbolic Memory}}.
\newblock   (\bibinfo{year}{2023}).
\newblock
\showeprint[arxiv]{cs.AI/2306.03901}


\bibitem[\protect\citeauthoryear{Johnson, Douze, and J{\'e}gou}{Johnson et~al\mbox{.}}{2019}]%
        {johnson2019billion}
\bibfield{author}{\bibinfo{person}{Jeff Johnson}, \bibinfo{person}{Matthijs Douze}, {and} \bibinfo{person}{Herv{\'e} J{\'e}gou}.} \bibinfo{year}{2019}\natexlab{}.
\newblock \showarticletitle{{Billion-scale similarity search with {GPUs}}}.
\newblock \bibinfo{journal}{{\em {IEEE Transactions on Big Data}\/}} \bibinfo{volume}{7}, \bibinfo{number}{3} (\bibinfo{year}{2019}), \bibinfo{pages}{535--547}.
\newblock


\bibitem[\protect\citeauthoryear{Joshi, Chen, Liu, Weld, Zettlemoyer, and Levy}{Joshi et~al\mbox{.}}{2020}]%
        {joshi2020spanbert}
\bibfield{author}{\bibinfo{person}{Mandar Joshi}, \bibinfo{person}{Danqi Chen}, \bibinfo{person}{Yinhan Liu}, \bibinfo{person}{Daniel~S Weld}, \bibinfo{person}{Luke Zettlemoyer}, {and} \bibinfo{person}{Omer Levy}.} \bibinfo{year}{2020}\natexlab{}.
\newblock \showarticletitle{{Spanbert: Improving pre-training by representing and predicting spans}}.
\newblock \bibinfo{journal}{{\em {Transactions of the association for computational linguistics}\/}}  \bibinfo{volume}{8} (\bibinfo{year}{2020}), \bibinfo{pages}{64--77}.
\newblock


\bibitem[\protect\citeauthoryear{Joulin, Grave, Bojanowski, Douze, J{\'e}gou, and Mikolov}{Joulin et~al\mbox{.}}{2016}]%
        {joulin2016fasttext}
\bibfield{author}{\bibinfo{person}{Armand Joulin}, \bibinfo{person}{Edouard Grave}, \bibinfo{person}{Piotr Bojanowski}, \bibinfo{person}{Matthijs Douze}, \bibinfo{person}{H{\'e}rve J{\'e}gou}, {and} \bibinfo{person}{Tomas Mikolov}.} \bibinfo{year}{2016}\natexlab{}.
\newblock \showarticletitle{{FastText.zip: Compressing text classification models}}.
\newblock \bibinfo{journal}{{\em {arXiv preprint arXiv:1612.03651}\/}} (\bibinfo{year}{2016}).
\newblock


\bibitem[\protect\citeauthoryear{Kamoonpuri and Sengar}{Kamoonpuri and Sengar}{2023}]%
        {kamoonpuri2023hi}
\bibfield{author}{\bibinfo{person}{Sana~Zehra Kamoonpuri} {and} \bibinfo{person}{Anita Sengar}.} \bibinfo{year}{2023}\natexlab{}.
\newblock \showarticletitle{{Hi, May AI help you? An analysis of the barriers impeding the implementation and use of artificial intelligence-enabled virtual assistants in retail}}.
\newblock \bibinfo{journal}{{\em Journal of Retailing and Consumer Services\/}}  \bibinfo{volume}{72} (\bibinfo{year}{2023}).
\newblock


\bibitem[\protect\citeauthoryear{Laban, Schnabel, Bennett, and Hearst}{Laban et~al\mbox{.}}{2022}]%
        {laban2022summac}
\bibfield{author}{\bibinfo{person}{Philippe Laban}, \bibinfo{person}{Tobias Schnabel}, \bibinfo{person}{Paul~N Bennett}, {and} \bibinfo{person}{Marti~A Hearst}.} \bibinfo{year}{2022}\natexlab{}.
\newblock \showarticletitle{{SummaC: Re-visiting NLI-based models for inconsistency detection in summarization}}.
\newblock \bibinfo{journal}{{\em {Transactions of the Association for Computational Linguistics}\/}}  \bibinfo{volume}{10} (\bibinfo{year}{2022}), \bibinfo{pages}{163--177}.
\newblock


\bibitem[\protect\citeauthoryear{Lan, Chen, Goodman, Gimpel, Sharma, and Soricut}{Lan et~al\mbox{.}}{2019}]%
        {DBLP:journals/corr/abs-1909-11942}
\bibfield{author}{\bibinfo{person}{Zhenzhong Lan}, \bibinfo{person}{Mingda Chen}, \bibinfo{person}{Sebastian Goodman}, \bibinfo{person}{Kevin Gimpel}, \bibinfo{person}{Piyush Sharma}, {and} \bibinfo{person}{Radu Soricut}.} \bibinfo{year}{2019}\natexlab{}.
\newblock \showarticletitle{{{ALBERT:} {A} Lite {BERT} for Self-supervised Learning of Language Representations}}.
\newblock \bibinfo{journal}{{\em {CoRR}\/}}  \bibinfo{volume}{abs/1909.11942} (\bibinfo{year}{2019}).
\newblock
\showeprint[arxiv]{1909.11942}
\showURL{%
\url{http://arxiv.org/abs/1909.11942}}


\bibitem[\protect\citeauthoryear{Lewis and Jones}{Lewis and Jones}{1996}]%
        {lewis1996natural}
\bibfield{author}{\bibinfo{person}{David~D Lewis} {and} \bibinfo{person}{Karen~Sp{\"a}rck Jones}.} \bibinfo{year}{1996}\natexlab{}.
\newblock \showarticletitle{{Natural language processing for information retrieval}}.
\newblock \bibinfo{journal}{{\em {Communications of the ACM}\/}} \bibinfo{volume}{39}, \bibinfo{number}{1} (\bibinfo{year}{1996}), \bibinfo{pages}{92--101}.
\newblock


\bibitem[\protect\citeauthoryear{Li, Deng, Liu, Wang, Li, Zhang, Liu, Xu, Xu, and Wang}{Li et~al\mbox{.}}{2024}]%
        {li2024digger}
\bibfield{author}{\bibinfo{person}{Haodong Li}, \bibinfo{person}{Gelei Deng}, \bibinfo{person}{Yi Liu}, \bibinfo{person}{Kailong Wang}, \bibinfo{person}{Yuekang Li}, \bibinfo{person}{Tianwei Zhang}, \bibinfo{person}{Yang Liu}, \bibinfo{person}{Guoai Xu}, \bibinfo{person}{Guosheng Xu}, {and} \bibinfo{person}{Haoyu Wang}.} \bibinfo{year}{2024}\natexlab{}.
\newblock \bibinfo{title}{{Digger: Detecting Copyright Content Mis-usage in Large Language Model Training}}.
\newblock   (\bibinfo{year}{2024}).
\newblock
\showeprint[arxiv]{cs.CR/2401.00676}


\bibitem[\protect\citeauthoryear{Li, Ji, Du, Li, and Wang}{Li et~al\mbox{.}}{2019}]%
        {LiJDLW19}
\bibfield{author}{\bibinfo{person}{Jinfeng Li}, \bibinfo{person}{Shouling Ji}, \bibinfo{person}{Tianyu Du}, \bibinfo{person}{Bo Li}, {and} \bibinfo{person}{Ting Wang}.} \bibinfo{year}{2019}\natexlab{}.
\newblock \showarticletitle{{TextBugger: Generating Adversarial Text Against Real-world Applications}}. In \bibinfo{booktitle}{{\em {NDSS}}}.
\newblock


\bibitem[\protect\citeauthoryear{Li, Liu, Dong, Zhao, Xue, Zhu, and Lu}{Li et~al\mbox{.}}{2021}]%
        {li2021hidden}
\bibfield{author}{\bibinfo{person}{Shaofeng Li}, \bibinfo{person}{Hui Liu}, \bibinfo{person}{Tian Dong}, \bibinfo{person}{Benjamin Zi~Hao Zhao}, \bibinfo{person}{Minhui Xue}, \bibinfo{person}{Haojin Zhu}, {and} \bibinfo{person}{Jialiang Lu}.} \bibinfo{year}{2021}\natexlab{}.
\newblock \bibinfo{title}{{Hidden Backdoors in Human-Centric Language Models}}.
\newblock   (\bibinfo{year}{2021}).
\newblock
\showeprint[arxiv]{cs.CL/2105.00164}


\bibitem[\protect\citeauthoryear{Liu}{Liu}{2022}]%
        {Liu_LlamaIndex_2022}
\bibfield{author}{\bibinfo{person}{Jerry Liu}.} \bibinfo{year}{2022}\natexlab{}.
\newblock \bibinfo{title}{{LlamaIndex}}.
\newblock   (\bibinfo{date}{11} \bibinfo{year}{2022}).
\newblock
\showDOI{%
\url{https://doi.org/10.5281/zenodo.1234}}


\bibitem[\protect\citeauthoryear{Liu, Deng, Li, Wang, Zhang, Liu, Wang, Zheng, and Liu}{Liu et~al\mbox{.}}{2023}]%
        {liu2023prompt}
\bibfield{author}{\bibinfo{person}{Yi Liu}, \bibinfo{person}{Gelei Deng}, \bibinfo{person}{Yuekang Li}, \bibinfo{person}{Kailong Wang}, \bibinfo{person}{Tianwei Zhang}, \bibinfo{person}{Yepang Liu}, \bibinfo{person}{Haoyu Wang}, \bibinfo{person}{Yan Zheng}, {and} \bibinfo{person}{Yang Liu}.} \bibinfo{year}{2023}\natexlab{}.
\newblock \showarticletitle{{Prompt Injection attack against LLM-integrated Applications}}.
\newblock \bibinfo{journal}{{\em {arXiv preprint}\/}} (\bibinfo{year}{2023}).
\newblock


\bibitem[\protect\citeauthoryear{Liu, Ott, Goyal, Du, Joshi, Chen, Levy, Lewis, Zettlemoyer, and Stoyanov}{Liu et~al\mbox{.}}{2019}]%
        {liu2019roberta}
\bibfield{author}{\bibinfo{person}{Yinhan Liu}, \bibinfo{person}{Myle Ott}, \bibinfo{person}{Naman Goyal}, \bibinfo{person}{Jingfei Du}, \bibinfo{person}{Mandar Joshi}, \bibinfo{person}{Danqi Chen}, \bibinfo{person}{Omer Levy}, \bibinfo{person}{Mike Lewis}, \bibinfo{person}{Luke Zettlemoyer}, {and} \bibinfo{person}{Veselin Stoyanov}.} \bibinfo{year}{2019}\natexlab{}.
\newblock \showarticletitle{{Roberta: A robustly optimized bert pretraining approach}}.
\newblock \bibinfo{journal}{{\em {arXiv preprint arXiv:1907.11692}\/}} (\bibinfo{year}{2019}).
\newblock


\bibitem[\protect\citeauthoryear{Loper and Bird}{Loper and Bird}{2002}]%
        {loper2002nltk}
\bibfield{author}{\bibinfo{person}{Edward Loper} {and} \bibinfo{person}{Steven Bird}.} \bibinfo{year}{2002}\natexlab{}.
\newblock \showarticletitle{{Nltk: The natural language toolkit}}.
\newblock \bibinfo{journal}{{\em {arXiv preprint cs/0205028}\/}} (\bibinfo{year}{2002}).
\newblock


\bibitem[\protect\citeauthoryear{Luo, Xie, and Ananiadou}{Luo et~al\mbox{.}}{2023}]%
        {luo2023chatgpt}
\bibfield{author}{\bibinfo{person}{Zheheng Luo}, \bibinfo{person}{Qianqian Xie}, {and} \bibinfo{person}{Sophia Ananiadou}.} \bibinfo{year}{2023}\natexlab{}.
\newblock \showarticletitle{{Chatgpt as a factual inconsistency evaluator for abstractive text summarization}}.
\newblock \bibinfo{journal}{{\em {arXiv preprint}\/}} (\bibinfo{year}{2023}).
\newblock


\bibitem[\protect\citeauthoryear{Ma, Wang, and Liu}{Ma et~al\mbox{.}}{2020}]%
        {Ma2020MetamorphicTA}
\bibfield{author}{\bibinfo{person}{Pingchuan Ma}, \bibinfo{person}{Shuai Wang}, {and} \bibinfo{person}{Jin Liu}.} \bibinfo{year}{2020}\natexlab{}.
\newblock \showarticletitle{{Metamorphic Testing and Certified Mitigation of Fairness Violations in NLP Models}}. In \bibinfo{booktitle}{{\em {International Joint Conference on Artificial Intelligence}}}.
\newblock
\showURL{%
\url{https://api.semanticscholar.org/CorpusID:220483049}}


\bibitem[\protect\citeauthoryear{Malkov and Yashunin}{Malkov and Yashunin}{2018}]%
        {malkov2018efficient}
\bibfield{author}{\bibinfo{person}{Yu~A Malkov} {and} \bibinfo{person}{Dmitry~A Yashunin}.} \bibinfo{year}{2018}\natexlab{}.
\newblock \showarticletitle{{Efficient and robust approximate nearest neighbor search using hierarchical navigable small world graphs}}.
\newblock \bibinfo{journal}{{\em {IEEE transactions on pattern analysis and machine intelligence}\/}} \bibinfo{volume}{42}, \bibinfo{number}{4} (\bibinfo{year}{2018}), \bibinfo{pages}{824--836}.
\newblock


\bibitem[\protect\citeauthoryear{{Michihiro Yasunaga and Jure Leskovec and Percy Liang}}{{Michihiro Yasunaga and Jure Leskovec and Percy Liang}}{2022}]%
        {yasunaga2022linkbert}
\bibfield{author}{\bibinfo{person}{{Michihiro Yasunaga and Jure Leskovec and Percy Liang}}.} \bibinfo{year}{2022}\natexlab{}.
\newblock \showarticletitle{{LinkBERT: Pretraining Language Models with Document Links}}. In \bibinfo{booktitle}{{\em {ACL}}}.
\newblock


\bibitem[\protect\citeauthoryear{microsoft}{microsoft}{2023}]%
        {mpnet}
\bibfield{author}{\bibinfo{person}{microsoft}.} \bibinfo{year}{2023}\natexlab{}.
\newblock \bibinfo{title}{{MPNet}}.
\newblock \bibinfo{howpublished}{\url{https://huggingface.co/microsoft/mpnet-base}}.   (\bibinfo{year}{2023}).
\newblock


\bibitem[\protect\citeauthoryear{Neelakantan, Xu, Puri, Radford, Han, Tworek, Yuan, Tezak, Kim, Hallacy, et~al\mbox{.}}{Neelakantan et~al\mbox{.}}{2022}]%
        {neelakantan2022text}
\bibfield{author}{\bibinfo{person}{Arvind Neelakantan}, \bibinfo{person}{Tao Xu}, \bibinfo{person}{Raul Puri}, \bibinfo{person}{Alec Radford}, \bibinfo{person}{Jesse~Michael Han}, \bibinfo{person}{Jerry Tworek}, \bibinfo{person}{Qiming Yuan}, \bibinfo{person}{Nikolas Tezak}, \bibinfo{person}{Jong~Wook Kim}, \bibinfo{person}{Chris Hallacy}, {et~al\mbox{.}}} \bibinfo{year}{2022}\natexlab{}.
\newblock \showarticletitle{{Text and code embeddings by contrastive pre-training}}.
\newblock \bibinfo{journal}{{\em {arXiv preprint arXiv:2201.10005}\/}} (\bibinfo{year}{2022}).
\newblock


\bibitem[\protect\citeauthoryear{Ni, {\'A}brego, Constant, Ma, Hall, Cer, and Yang}{Ni et~al\mbox{.}}{2021a}]%
        {ni2021sentence}
\bibfield{author}{\bibinfo{person}{Jianmo Ni}, \bibinfo{person}{Gustavo~Hern{\'a}ndez {\'A}brego}, \bibinfo{person}{Noah Constant}, \bibinfo{person}{Ji Ma}, \bibinfo{person}{Keith~B Hall}, \bibinfo{person}{Daniel Cer}, {and} \bibinfo{person}{Yinfei Yang}.} \bibinfo{year}{2021}\natexlab{a}.
\newblock \showarticletitle{{Sentence-t5: Scalable sentence encoders from pre-trained text-to-text models}}.
\newblock \bibinfo{journal}{{\em {arXiv preprint arXiv:2108.08877}\/}} (\bibinfo{year}{2021}).
\newblock


\bibitem[\protect\citeauthoryear{Ni, Qu, Lu, Dai, {\'A}brego, Ma, Zhao, Luan, Hall, Chang, et~al\mbox{.}}{Ni et~al\mbox{.}}{2021b}]%
        {ni2021large}
\bibfield{author}{\bibinfo{person}{Jianmo Ni}, \bibinfo{person}{Chen Qu}, \bibinfo{person}{Jing Lu}, \bibinfo{person}{Zhuyun Dai}, \bibinfo{person}{Gustavo~Hern{\'a}ndez {\'A}brego}, \bibinfo{person}{Ji Ma}, \bibinfo{person}{Vincent~Y Zhao}, \bibinfo{person}{Yi Luan}, \bibinfo{person}{Keith~B Hall}, \bibinfo{person}{Ming-Wei Chang}, {et~al\mbox{.}}} \bibinfo{year}{2021}\natexlab{b}.
\newblock \showarticletitle{{Large dual encoders are generalizable retrievers}}.
\newblock \bibinfo{journal}{{\em {arXiv preprint arXiv:2112.07899}\/}} (\bibinfo{year}{2021}).
\newblock


\bibitem[\protect\citeauthoryear{Penedo, Malartic, Hesslow, Cojocaru, Cappelli, Alobeidli, Pannier, Almazrouei, and Launay}{Penedo et~al\mbox{.}}{2023}]%
        {refinedweb}
\bibfield{author}{\bibinfo{person}{Guilherme Penedo}, \bibinfo{person}{Quentin Malartic}, \bibinfo{person}{Daniel Hesslow}, \bibinfo{person}{Ruxandra Cojocaru}, \bibinfo{person}{Alessandro Cappelli}, \bibinfo{person}{Hamza Alobeidli}, \bibinfo{person}{Baptiste Pannier}, \bibinfo{person}{Ebtesam Almazrouei}, {and} \bibinfo{person}{Julien Launay}.} \bibinfo{year}{2023}\natexlab{}.
\newblock \showarticletitle{{The {R}efined{W}eb dataset for {F}alcon {LLM}: outperforming curated corpora with web data, and web data only}}.
\newblock \bibinfo{journal}{{\em {arXiv preprint arXiv:2306.01116}\/}} (\bibinfo{year}{2023}).
\newblock
\showeprint{2306.01116}
\showURL{%
\url{https://arxiv.org/abs/2306.01116}}


\bibitem[\protect\citeauthoryear{Perez and Ribeiro}{Perez and Ribeiro}{2022}]%
        {perez2022ignore}
\bibfield{author}{\bibinfo{person}{Fábio Perez} {and} \bibinfo{person}{Ian Ribeiro}.} \bibinfo{year}{2022}\natexlab{}.
\newblock \bibinfo{title}{{Ignore Previous Prompt: Attack Techniques For Language Models}}.
\newblock   (\bibinfo{year}{2022}).
\newblock
\showeprint[{arXiv}]{cs.CL/2211.09527}


\bibitem[\protect\citeauthoryear{pgvector}{pgvector}{2023}]%
        {pgvector}
\bibfield{author}{\bibinfo{person}{pgvector}.} \bibinfo{year}{2023}\natexlab{}.
\newblock \bibinfo{title}{{PGVector}}.
\newblock \bibinfo{howpublished}{\url{https://github.com/pgvector/pgvector}}.   (\bibinfo{year}{2023}).
\newblock


\bibitem[\protect\citeauthoryear{pinecone}{pinecone}{2023}]%
        {pinecone}
\bibfield{author}{\bibinfo{person}{pinecone}.} \bibinfo{year}{2023}\natexlab{}.
\newblock \bibinfo{title}{{Pinecone}}.
\newblock \bibinfo{howpublished}{\url{https://www.pinecone.io/}}.   (\bibinfo{year}{2023}).
\newblock


\bibitem[\protect\citeauthoryear{Ribeiro, Wu, Guestrin, and Singh}{Ribeiro et~al\mbox{.}}{2020}]%
        {ribeiro-etal-2020-beyond}
\bibfield{author}{\bibinfo{person}{Marco~Tulio Ribeiro}, \bibinfo{person}{Tongshuang Wu}, \bibinfo{person}{Carlos Guestrin}, {and} \bibinfo{person}{Sameer Singh}.} \bibinfo{year}{2020}\natexlab{}.
\newblock \showarticletitle{{Beyond Accuracy: Behavioral Testing of {NLP} Models with {C}heck{L}ist}}. In \bibinfo{booktitle}{{\em {ACL}}}. \bibinfo{pages}{4902--4912}.
\newblock


\bibitem[\protect\citeauthoryear{Sait and Ishak}{Sait and Ishak}{2023}]%
        {sait2023deep}
\bibfield{author}{\bibinfo{person}{Abdul Rahaman~Wahab Sait} {and} \bibinfo{person}{Mohamad~Khairi Ishak}.} \bibinfo{year}{2023}\natexlab{}.
\newblock \showarticletitle{{Deep learning with natural language processing enabled sentimental analysis on sarcasm classification}}.
\newblock \bibinfo{journal}{{\em {Comput. Syst. Sci. Eng}\/}} \bibinfo{volume}{44}, \bibinfo{number}{3} (\bibinfo{year}{2023}), \bibinfo{pages}{2553--2567}.
\newblock


\bibitem[\protect\citeauthoryear{Sanh, Debut, Chaumond, and Wolf}{Sanh et~al\mbox{.}}{2019}]%
        {Sanh2019DistilBERTAD}
\bibfield{author}{\bibinfo{person}{Victor Sanh}, \bibinfo{person}{Lysandre Debut}, \bibinfo{person}{Julien Chaumond}, {and} \bibinfo{person}{Thomas Wolf}.} \bibinfo{year}{2019}\natexlab{}.
\newblock \showarticletitle{{DistilBERT, a distilled version of BERT: smaller, faster, cheaper and lighter}}.
\newblock \bibinfo{journal}{{\em {ArXiv}\/}}  \bibinfo{volume}{abs/1910.01108} (\bibinfo{year}{2019}).
\newblock


\bibitem[\protect\citeauthoryear{Santos, da~Silveira, de~Andrade, Delamaro, and do~Rocio Senger~de Souza}{Santos et~al\mbox{.}}{2020}]%
        {Santos2020AnES}
\bibfield{author}{\bibinfo{person}{Sebasti{\~a}o Santos}, \bibinfo{person}{Beatriz Nogueira~Carvalho da Silveira}, \bibinfo{person}{Stev{\~a}o~Alves de Andrade}, \bibinfo{person}{M{\'a}rcio~Eduardo Delamaro}, {and} \bibinfo{person}{Simone do~Rocio Senger~de Souza}.} \bibinfo{year}{2020}\natexlab{}.
\newblock \showarticletitle{{An Experimental Study on Applying Metamorphic Testing in Machine Learning Applications}}.
\newblock \bibinfo{journal}{{\em {Proceedings of the 5th Brazilian Symposium on Systematic and Automated Software Testing}\/}} (\bibinfo{year}{2020}).
\newblock
\showURL{%
\url{https://api.semanticscholar.org/CorpusID:225040791}}


\bibitem[\protect\citeauthoryear{Schuster, Fisch, and Barzilay}{Schuster et~al\mbox{.}}{2021}]%
        {schuster2021get}
\bibfield{author}{\bibinfo{person}{Tal Schuster}, \bibinfo{person}{Adam Fisch}, {and} \bibinfo{person}{Regina Barzilay}.} \bibinfo{year}{2021}\natexlab{}.
\newblock \showarticletitle{{Get your vitamin C! robust fact verification with contrastive evidence}}.
\newblock \bibinfo{journal}{{\em {arXiv preprint arXiv:2103.08541}\/}} (\bibinfo{year}{2021}).
\newblock


\bibitem[\protect\citeauthoryear{Scialom, Dray, Gallinari, Lamprier, Piwowarski, Staiano, and Wang}{Scialom et~al\mbox{.}}{2021}]%
        {scialom2021questeval}
\bibfield{author}{\bibinfo{person}{Thomas Scialom}, \bibinfo{person}{Paul-Alexis Dray}, \bibinfo{person}{Patrick Gallinari}, \bibinfo{person}{Sylvain Lamprier}, \bibinfo{person}{Benjamin Piwowarski}, \bibinfo{person}{Jacopo Staiano}, {and} \bibinfo{person}{Alex Wang}.} \bibinfo{year}{2021}\natexlab{}.
\newblock \showarticletitle{{Questeval: Summarization asks for fact-based evaluation}}.
\newblock \bibinfo{journal}{{\em {arXiv preprint arXiv:2103.12693}\/}} (\bibinfo{year}{2021}).
\newblock


\bibitem[\protect\citeauthoryear{scipy}{scipy}{2023}]%
        {scipy}
\bibfield{author}{\bibinfo{person}{scipy}.} \bibinfo{year}{2023}\natexlab{}.
\newblock \bibinfo{title}{{Fundamental algorithms for scientific computing in Python}}.
\newblock \bibinfo{howpublished}{\url{https://scipy.org/}}.   (\bibinfo{year}{2023}).
\newblock


\bibitem[\protect\citeauthoryear{Spotify}{Spotify}{2023}]%
        {annoy}
\bibfield{author}{\bibinfo{person}{Spotify}.} \bibinfo{year}{2023}\natexlab{}.
\newblock \bibinfo{title}{{Annoy}}.
\newblock \bibinfo{howpublished}{\url{https://github.com/spotify/annoy?tab=readme-ov-file}}.   (\bibinfo{year}{2023}).
\newblock


\bibitem[\protect\citeauthoryear{Sun, Zhang, Xiong, Harman, Papadakis, and Zhang}{Sun et~al\mbox{.}}{2022}]%
        {Sun2022ImprovingMT}
\bibfield{author}{\bibinfo{person}{Zeyu Sun}, \bibinfo{person}{J Zhang}, \bibinfo{person}{Yingfei Xiong}, \bibinfo{person}{Mark Harman}, \bibinfo{person}{Mike Papadakis}, {and} \bibinfo{person}{Lu Zhang}.} \bibinfo{year}{2022}\natexlab{}.
\newblock \showarticletitle{{Improving Machine Translation Systems via Isotopic Replacement}}. In \bibinfo{booktitle}{{\em {ICSE}}}. \bibinfo{pages}{1181--1192}.
\newblock


\bibitem[\protect\citeauthoryear{Tan, Joty, Baxter, Taeihagh, Bennett, and Kan}{Tan et~al\mbox{.}}{2021}]%
        {tan2021reliability}
\bibfield{author}{\bibinfo{person}{Samson Tan}, \bibinfo{person}{Shafiq Joty}, \bibinfo{person}{Kathy Baxter}, \bibinfo{person}{Araz Taeihagh}, \bibinfo{person}{Gregory~A. Bennett}, {and} \bibinfo{person}{Min-Yen Kan}.} \bibinfo{year}{2021}\natexlab{}.
\newblock \bibinfo{title}{{Reliability Testing for Natural Language Processing Systems}}.
\newblock   (\bibinfo{year}{2021}).
\newblock
\showeprint[arxiv]{cs.LG/2105.02590}


\bibitem[\protect\citeauthoryear{Touvron, Lavril, Izacard, Martinet, Lachaux, Lacroix, Rozi{\`e}re, Goyal, Hambro, Azhar, et~al\mbox{.}}{Touvron et~al\mbox{.}}{2023a}]%
        {touvron2023llama}
\bibfield{author}{\bibinfo{person}{Hugo Touvron}, \bibinfo{person}{Thibaut Lavril}, \bibinfo{person}{Gautier Izacard}, \bibinfo{person}{Xavier Martinet}, \bibinfo{person}{Marie-Anne Lachaux}, \bibinfo{person}{Timoth{\'e}e Lacroix}, \bibinfo{person}{Baptiste Rozi{\`e}re}, \bibinfo{person}{Naman Goyal}, \bibinfo{person}{Eric Hambro}, \bibinfo{person}{Faisal Azhar}, {et~al\mbox{.}}} \bibinfo{year}{2023}\natexlab{a}.
\newblock \showarticletitle{{Llama: Open and efficient foundation language models}}.
\newblock \bibinfo{journal}{{\em {arXiv preprint arXiv:2302.13971}\/}} (\bibinfo{year}{2023}).
\newblock


\bibitem[\protect\citeauthoryear{Touvron, Martin, Stone, Albert, Almahairi, Babaei, Bashlykov, Batra, Bhargava, Bhosale, et~al\mbox{.}}{Touvron et~al\mbox{.}}{2023b}]%
        {touvron2023llama2}
\bibfield{author}{\bibinfo{person}{Hugo Touvron}, \bibinfo{person}{Louis Martin}, \bibinfo{person}{Kevin Stone}, \bibinfo{person}{Peter Albert}, \bibinfo{person}{Amjad Almahairi}, \bibinfo{person}{Yasmine Babaei}, \bibinfo{person}{Nikolay Bashlykov}, \bibinfo{person}{Soumya Batra}, \bibinfo{person}{Prajjwal Bhargava}, \bibinfo{person}{Shruti Bhosale}, {et~al\mbox{.}}} \bibinfo{year}{2023}\natexlab{b}.
\newblock \showarticletitle{{Llama 2: Open foundation and fine-tuned chat models}}.
\newblock \bibinfo{journal}{{\em {arXiv preprint arXiv:2307.09288}\/}} (\bibinfo{year}{2023}).
\newblock


\bibitem[\protect\citeauthoryear{unum cloud}{unum cloud}{2023}]%
        {uform}
\bibfield{author}{\bibinfo{person}{unum cloud}.} \bibinfo{year}{2023}\natexlab{}.
\newblock \bibinfo{title}{{Uform}}.
\newblock \bibinfo{howpublished}{\url{https://huggingface.co/unum-cloud/uform-vl-english}}.   (\bibinfo{year}{2023}).
\newblock


\bibitem[\protect\citeauthoryear{Voorhees}{Voorhees}{1999}]%
        {voorhees1999natural}
\bibfield{author}{\bibinfo{person}{Ellen~M Voorhees}.} \bibinfo{year}{1999}\natexlab{}.
\newblock \showarticletitle{{Natural language processing and information retrieval}}.
\newblock In \bibinfo{booktitle}{{\em International summer school on information extraction}}. \bibinfo{pages}{32--48}.
\newblock


\bibitem[\protect\citeauthoryear{Wang, Yi, Guo, Jin, Xu, Li, Wang, Guo, Li, Xu, et~al\mbox{.}}{Wang et~al\mbox{.}}{2021}]%
        {2021milvus}
\bibfield{author}{\bibinfo{person}{Jianguo Wang}, \bibinfo{person}{Xiaomeng Yi}, \bibinfo{person}{Rentong Guo}, \bibinfo{person}{Hai Jin}, \bibinfo{person}{Peng Xu}, \bibinfo{person}{Shengjun Li}, \bibinfo{person}{Xiangyu Wang}, \bibinfo{person}{Xiangzhou Guo}, \bibinfo{person}{Chengming Li}, \bibinfo{person}{Xiaohai Xu}, {et~al\mbox{.}}} \bibinfo{year}{2021}\natexlab{}.
\newblock \showarticletitle{{Milvus: A Purpose-Built Vector Data Management System}}. In \bibinfo{booktitle}{{\em {Proceedings of the 2021 International Conference on Management of Data}}}. \bibinfo{pages}{2614--2627}.
\newblock


\bibitem[\protect\citeauthoryear{Wang, Sun, Xiang, Wu, Ding, Gong, Feng, Shang, Zhao, Pang, et~al\mbox{.}}{Wang et~al\mbox{.}}{2021}]%
        {wang2021ernie}
\bibfield{author}{\bibinfo{person}{Shuohuan Wang}, \bibinfo{person}{Yu Sun}, \bibinfo{person}{Yang Xiang}, \bibinfo{person}{Zhihua Wu}, \bibinfo{person}{Siyu Ding}, \bibinfo{person}{Weibao Gong}, \bibinfo{person}{Shikun Feng}, \bibinfo{person}{Junyuan Shang}, \bibinfo{person}{Yanbin Zhao}, \bibinfo{person}{Chao Pang}, {et~al\mbox{.}}} \bibinfo{year}{2021}\natexlab{}.
\newblock \showarticletitle{{Ernie 3.0 titan: Exploring larger-scale knowledge enhanced pre-training for language understanding and generation}}.
\newblock \bibinfo{journal}{{\em {arXiv preprint arXiv:2112.12731}\/}} (\bibinfo{year}{2021}).
\newblock


\bibitem[\protect\citeauthoryear{Wang, Huang, Wu, Zhang, Huang, Li, He, and Lyu}{Wang et~al\mbox{.}}{2023}]%
        {Wang_ICSE23}
\bibfield{author}{\bibinfo{person}{Wenxuan Wang}, \bibinfo{person}{Jen-tse Huang}, \bibinfo{person}{Weibin Wu}, \bibinfo{person}{Jianping Zhang}, \bibinfo{person}{Yizhan Huang}, \bibinfo{person}{Shuqing Li}, \bibinfo{person}{Pinjia He}, {and} \bibinfo{person}{Michael~R. Lyu}.} \bibinfo{year}{2023}\natexlab{}.
\newblock \showarticletitle{{MTTM: Metamorphic Testing for Textual Content Moderation Software}}. In \bibinfo{booktitle}{{\em {ICSE}}}. \bibinfo{pages}{2387--2399}.
\newblock


\bibitem[\protect\citeauthoryear{Wang, Wei, Dong, Bao, Yang, and Zhou}{Wang et~al\mbox{.}}{2020}]%
        {wang2020minilm}
\bibfield{author}{\bibinfo{person}{Wenhui Wang}, \bibinfo{person}{Furu Wei}, \bibinfo{person}{Li Dong}, \bibinfo{person}{Hangbo Bao}, \bibinfo{person}{Nan Yang}, {and} \bibinfo{person}{Ming Zhou}.} \bibinfo{year}{2020}\natexlab{}.
\newblock \showarticletitle{{Minilm: Deep self-attention distillation for task-agnostic compression of pre-trained transformers}}.
\newblock \bibinfo{journal}{{\em {Advances in Neural Information Processing Systems}\/}}  \bibinfo{volume}{33} (\bibinfo{year}{2020}), \bibinfo{pages}{5776--5788}.
\newblock


\bibitem[\protect\citeauthoryear{Wei, Wang, Schuurmans, Bosma, Xia, Chi, Le, Zhou, et~al\mbox{.}}{Wei et~al\mbox{.}}{2022}]%
        {wei2022chain}
\bibfield{author}{\bibinfo{person}{Jason Wei}, \bibinfo{person}{Xuezhi Wang}, \bibinfo{person}{Dale Schuurmans}, \bibinfo{person}{Maarten Bosma}, \bibinfo{person}{Fei Xia}, \bibinfo{person}{Ed Chi}, \bibinfo{person}{Quoc~V Le}, \bibinfo{person}{Denny Zhou}, {et~al\mbox{.}}} \bibinfo{year}{2022}\natexlab{}.
\newblock \showarticletitle{{Chain-of-thought prompting elicits reasoning in large language models}}.
\newblock \bibinfo{journal}{{\em {Advances in Neural Information Processing Systems}\/}}  \bibinfo{volume}{35} (\bibinfo{year}{2022}), \bibinfo{pages}{24824--24837}.
\newblock


\bibitem[\protect\citeauthoryear{Weller, Lawrie, and Van~Durme}{Weller et~al\mbox{.}}{2023}]%
        {weller2023nevir}
\bibfield{author}{\bibinfo{person}{Orion Weller}, \bibinfo{person}{Dawn Lawrie}, {and} \bibinfo{person}{Benjamin Van~Durme}.} \bibinfo{year}{2023}\natexlab{}.
\newblock \showarticletitle{{NevIR: Negation in Neural Information Retrieval}}.
\newblock \bibinfo{journal}{{\em {arXiv preprint arXiv:2305.07614}\/}} (\bibinfo{year}{2023}).
\newblock


\bibitem[\protect\citeauthoryear{White, Rastogi, Duh, and Van~Durme}{White et~al\mbox{.}}{2017}]%
        {white2017inference}
\bibfield{author}{\bibinfo{person}{Aaron~Steven White}, \bibinfo{person}{Pushpendre Rastogi}, \bibinfo{person}{Kevin Duh}, {and} \bibinfo{person}{Benjamin Van~Durme}.} \bibinfo{year}{2017}\natexlab{}.
\newblock \showarticletitle{{Inference is everything: Recasting semantic resources into a unified evaluation framework}}. In \bibinfo{booktitle}{{\em {Proceedings of the Eighth International Joint Conference on Natural Language Processing (Volume 1: Long Papers)}}}. \bibinfo{pages}{996--1005}.
\newblock


\bibitem[\protect\citeauthoryear{William, Shrivastava, Chauhan, Raja, Ojha, and Kumar}{William et~al\mbox{.}}{2023}]%
        {william2023natural}
\bibfield{author}{\bibinfo{person}{P William}, \bibinfo{person}{Anurag Shrivastava}, \bibinfo{person}{Premanand~S Chauhan}, \bibinfo{person}{Mudasir Raja}, \bibinfo{person}{Sudhir~Baijnath Ojha}, {and} \bibinfo{person}{Keshav Kumar}.} \bibinfo{year}{2023}\natexlab{}.
\newblock \showarticletitle{{Natural Language processing implementation for sentiment analysis on tweets}}.
\newblock In \bibinfo{booktitle}{{\em {MRCN}}}. \bibinfo{pages}{317--327}.
\newblock


\bibitem[\protect\citeauthoryear{Xiao, Liu, Yuan, Pang, and Wang}{Xiao et~al\mbox{.}}{2022}]%
        {Xiao2022MetamorphicTO}
\bibfield{author}{\bibinfo{person}{Dongwei Xiao}, \bibinfo{person}{Zhibo Liu}, \bibinfo{person}{Yuanyuan Yuan}, \bibinfo{person}{Qi Pang}, {and} \bibinfo{person}{Shuai Wang}.} \bibinfo{year}{2022}\natexlab{}.
\newblock \showarticletitle{{Metamorphic Testing of Deep Learning Compilers}}.
\newblock \bibinfo{journal}{{\em {Proceedings of the ACM on Measurement and Analysis of Computing Systems}\/}}  \bibinfo{volume}{6} (\bibinfo{year}{2022}), \bibinfo{pages}{1 -- 28}.
\newblock
\showURL{%
\url{https://api.semanticscholar.org/CorpusID:247159402}}


\bibitem[\protect\citeauthoryear{Xiao, Xiao, Dong, Ji, and Zhang}{Xiao et~al\mbox{.}}{2023}]%
        {xiao2023leap}
\bibfield{author}{\bibinfo{person}{Mingxuan Xiao}, \bibinfo{person}{Yan Xiao}, \bibinfo{person}{Hai Dong}, \bibinfo{person}{Shunhui Ji}, {and} \bibinfo{person}{Pengcheng Zhang}.} \bibinfo{year}{2023}\natexlab{}.
\newblock \showarticletitle{{LEAP: Efficient and Automated Test Method for NLP Software}}.
\newblock \bibinfo{journal}{{\em {arXiv preprint arXiv:2308.11284}\/}} (\bibinfo{year}{2023}).
\newblock


\bibitem[\protect\citeauthoryear{Yang, Li, Zhang, Chen, and Cheng}{Yang et~al\mbox{.}}{2023}]%
        {yang2023exploring}
\bibfield{author}{\bibinfo{person}{Xianjun Yang}, \bibinfo{person}{Yan Li}, \bibinfo{person}{Xinlu Zhang}, \bibinfo{person}{Haifeng Chen}, {and} \bibinfo{person}{Wei Cheng}.} \bibinfo{year}{2023}\natexlab{}.
\newblock \showarticletitle{{Exploring the limits of chatgpt for query or aspect-based text summarization}}.
\newblock \bibinfo{journal}{{\em {arXiv preprint}\/}} (\bibinfo{year}{2023}).
\newblock


\bibitem[\protect\citeauthoryear{Yang, Dai, Yang, Carbonell, Salakhutdinov, and Le}{Yang et~al\mbox{.}}{2019}]%
        {DBLP:journals/corr/abs-1906-08237}
\bibfield{author}{\bibinfo{person}{Zhilin Yang}, \bibinfo{person}{Zihang Dai}, \bibinfo{person}{Yiming Yang}, \bibinfo{person}{Jaime~G. Carbonell}, \bibinfo{person}{Ruslan Salakhutdinov}, {and} \bibinfo{person}{Quoc~V. Le}.} \bibinfo{year}{2019}\natexlab{}.
\newblock \showarticletitle{{XLNet: Generalized Autoregressive Pretraining for Language Understanding}}.
\newblock \bibinfo{journal}{{\em {CoRR}\/}}  \bibinfo{volume}{abs/1906.08237} (\bibinfo{year}{2019}).
\newblock
\showeprint{1906.08237}
\showURL{%
\url{http://arxiv.org/abs/1906.08237}}


\bibitem[\protect\citeauthoryear{Yew, Schraagen, Otte, and van Diessen}{Yew et~al\mbox{.}}{2023}]%
        {yew2023transforming}
\bibfield{author}{\bibinfo{person}{Arister~NJ Yew}, \bibinfo{person}{Marijn Schraagen}, \bibinfo{person}{Willem~M Otte}, {and} \bibinfo{person}{Eric van Diessen}.} \bibinfo{year}{2023}\natexlab{}.
\newblock \showarticletitle{{Transforming epilepsy research: A systematic review on natural language processing applications}}.
\newblock \bibinfo{journal}{{\em {Epilepsia}\/}} \bibinfo{volume}{64}, \bibinfo{number}{2} (\bibinfo{year}{2023}), \bibinfo{pages}{292--305}.
\newblock


\bibitem[\protect\citeauthoryear{Yuan, Pang, and Wang}{Yuan et~al\mbox{.}}{2022}]%
        {Yuan2022UnveilingHD}
\bibfield{author}{\bibinfo{person}{Yuanyuan Yuan}, \bibinfo{person}{Qi Pang}, {and} \bibinfo{person}{Shuai Wang}.} \bibinfo{year}{2022}\natexlab{}.
\newblock \showarticletitle{{Unveiling Hidden DNN Defects with Decision-Based Metamorphic Testing}}.
\newblock \bibinfo{journal}{{\em {Proceedings of the 37th IEEE/ACM International Conference on Automated Software Engineering}\/}} (\bibinfo{year}{2022}).
\newblock
\showURL{%
\url{https://api.semanticscholar.org/CorpusID:252816122}}


\bibitem[\protect\citeauthoryear{Zhang, Baldridge, and He}{Zhang et~al\mbox{.}}{2019}]%
        {zhang2019paws}
\bibfield{author}{\bibinfo{person}{Yuan Zhang}, \bibinfo{person}{Jason Baldridge}, {and} \bibinfo{person}{Luheng He}.} \bibinfo{year}{2019}\natexlab{}.
\newblock \showarticletitle{{PAWS: Paraphrase adversaries from word scrambling}}.
\newblock \bibinfo{journal}{{\em {arXiv preprint arXiv:1904.01130}\/}} (\bibinfo{year}{2019}).
\newblock


\bibitem[\protect\citeauthoryear{Zhou, Silvasstar, Clark, Salyers, Chavez, and Bull}{Zhou et~al\mbox{.}}{2023}]%
        {zhou2023artificially}
\bibfield{author}{\bibinfo{person}{Shuo Zhou}, \bibinfo{person}{Joshva Silvasstar}, \bibinfo{person}{Christopher Clark}, \bibinfo{person}{Adam~J Salyers}, \bibinfo{person}{Catia Chavez}, {and} \bibinfo{person}{Sheana~S Bull}.} \bibinfo{year}{2023}\natexlab{}.
\newblock \showarticletitle{{An artificially intelligent, natural language processing chatbot designed to promote COVID-19 vaccination: A proof-of-concept pilot study}}.
\newblock \bibinfo{journal}{{\em {Digital Health}\/}}  \bibinfo{volume}{9} (\bibinfo{year}{2023}).
\newblock


\bibitem[\protect\citeauthoryear{zilliztech}{zilliztech}{2023a}]%
        {gptcache}
\bibfield{author}{\bibinfo{person}{zilliztech}.} \bibinfo{year}{2023}\natexlab{a}.
\newblock \bibinfo{title}{{GPTCache}}.
\newblock \bibinfo{howpublished}{\url{https://github.com/zilliztech/GPTCache}}.   (\bibinfo{year}{2023}).
\newblock


\bibitem[\protect\citeauthoryear{zilliztech}{zilliztech}{2023b}]%
        {albertonnx}
\bibfield{author}{\bibinfo{person}{zilliztech}.} \bibinfo{year}{2023}\natexlab{b}.
\newblock \bibinfo{title}{{Paraphrase-albert-onnx}}.
\newblock \bibinfo{howpublished}{\url{https://huggingface.co/GPTCache/paraphrase-albert-onnx}}.   (\bibinfo{year}{2023}).
\newblock


\end{thebibliography}

\end{document}